\documentclass[review,1p,number,sort&compress]{elsarticle}

\usepackage{graphicx}
\usepackage[table,xcdraw]{xcolor}
\usepackage{float}
\usepackage{mathtools}
\usepackage{hyperref}
\usepackage{amsmath}
\usepackage{multirow}
\usepackage{comment}

\usepackage{lineno}

\begin{document}

\title{Millikelvin-precision temperature sensing for advanced cryogenic detectors}

\author[1]{M. Antonova}
\author[1]{J. Cap\'o}
\author[1]{A. Cervera}
\author[1]{P. Fern\'andez\fnref{fn2}}
\author[1]{M. \'A. Garc\'ia-Peris\corref{corr}\fnref{fn1}}
\ead{miguel.garciaperis@manchester.ac.uk}
\author[2]{X. Pons}

\affiliation[1]{Instituto de Fisica Corpuscular, Catedratico Jose Beltran, 2 E-46980 Paterna (Valencia), Spain}
\affiliation[2]{CERN, Espl. des Particules 1, 1211 Meyrin, Switzerland}

\cortext[corr]{Corresponding author}
\fntext[fn1]{now in the University of Manchester}
\fntext[fn2]{now at Donostia International Physics Center, DIPC}
\date{\today}

\begin{abstract}
Precise temperature monitoring  ---to the level of a few milli-Kelvin--- is essential for the operation of large-scale cryostats requiring a recirculation system. In particular, the performance of Liquid Argon Time Projection Chambers ---such as those planned for the DUNE experiment--- strongly relies on proper argon purification and mixing, which can be characterized by a sufficiently dense grid of high-precision temperature probes. In this article, we present a novel technique for the cross-calibration of Resistance Temperature Detectors in cryogenic liquids, developed as part of the temperature monitoring system for a DUNE prototype. This calibration has enabled the validation and optimization of the system’s components, achieving an unprecedented precision of 2.5 mK.

\end{abstract}


\maketitle

\section{Introduction}
\label{sec:introduction}

\noindent Precise temperature sensing is essential in large cryogenic detectors, where the circulation and purification of cryogenic liquids significantly impact overall detector performance. Uncontrolled temperature gradients in such systems can disrupt the cryostat dynamics and ultimately affect the experiment's physics sensitivity. In this work, we present the calibration procedure of a novel temperature monitoring system developed for ProtoDUNE Single-Phase (SP), a prototype of the future DUNE experiment. This procedure has been essential for ensuring accurate measurements and for validating and optimizing the various components of the system.

DUNE, which stands for Deep Underground Neutrino Experiment~\cite{dune_tdr1}, is expected to begin taking data toward the end of the decade. It aims to perform comprehensive neutrino oscillation analyses ---broadly exploring the Charge-Parity (CP) violation parameters phase-space and resolving the neutrino-mass hierarchy problem \cite{bib:dune_osc}. Its extense physics program also includes searches for hypothetical proton decay channels \cite{bib:dune_tdr2}, multi-messenger astronomy from supernovae and neutrino bursts \cite{bib:dune_supernova}, and explorations of Beyond the Standard Model (BSM) physics \cite{bib:dune_bsm}. Utilizing the most powerful neutrino beam ever constructed, generated at Fermilab, the experiment adopts a long-baseline neutrino oscillation approach with two detectors. The Near Detector, also at Fermilab, will characterize the unoscillated neutrino beam. The Far Detector (FD), located in the Sandford Underground Research Facility (SURF), 1300 km away from the ND and 1.5 km underground, will measure the oscillated flux.

In the Phase I, DUNE Far Detector will consist of two Liquid Argon Time Projection Chambers (LArTPCs) and is anticipated to begin operations by 2029. DUNE phase II will complement with two more modules by mid 30's, aiming at a total fiducial mass of 40 kilotonnes. The detector technology has been established, with excellent tracking and calorimetric capabilities, in several smaller-scale experiments \cite{icarus,microboone}. The first demonstration of the technology at the kilotonne scale has been carried out at the CERN Neutrino Platform as part of the ProtoDUNE program. In particular, the ProtoDUNE-SP experiment~\cite{pdsp_tdr} replicated the components of the DUNE FD Horizontal Drift (HD)~\cite{dune_tdr4} configuration at a scale 1:1, using a total argon mass 20 times smaller (770 tonnes). It was operated from mid-2018 to mid-2020, constituting the largest monolithic LArTPC to ever constructed and operated up to date~\cite{pdsp_1,pdsp_2}.


During ProtoDUNE-SP operation, the level of impurities was kept way below 100 ppt oxygen equivalent using a cryogenic recirculation and purification system~\cite{pdsp_tdr,pdsp_2}. This is critical in an LArTPC, as the 3D images generated by charged particles traversing the detector can be significantly degraded by impurities in the medium—such as nitrogen, oxygen, and water—which absorb ionization electrons, thereby deteriorating spatial resolution and introducing biases in energy measurements. Three purity monitors, based on the ICARUS design~\cite{PrMs}, were installed outside the active volume of the TPC to measure the electron lifetime, which is inversely related to the residual concentration of impurities. They ran twice a day to monitor the argon purity and thus to provide the necessary corrections for posterior data analysis.

Achieving the necessary argon purity has been possible thanks to the studies based on computational fluid dynamic (CFD) simulations, and the experience gained with previous LArTPC demonstrators such as LAPD~\cite{lapd} and the 35-tonne prototype~\cite{35t_1,35t_2}, which have paved the way to operate large-scale cryostats requiring low concentrations of impurities. These studies have also shown that it exists a strong correlation between temperature and purity in the liquid argon volume. The distribution of impurities is insensitive to small ($\mathcal{O}\sim$1 K) absolute temperature variations, but strongly depends on the relative vertical temperature gradient. Since the bulk volume in the cryostats must be continuously mixed with the incoming purified argon to ensure uniform purification of the entire LAr volume, the temperature distribution serves as a clear indicator of the mixing process: a homogeneous temperature distribution suggests proper mixing, while large temperature gradients signal inadequate mixing. If the LAr bulk volume is not mixed appropriately, a stratification regime can develop: a significant portion of the liquid remains unpurified, generating `dead' regions inside the detector. Thus, continuous monitoring of this temperature gradient can identify and mitigate potential failures of the purification system. In ProtoDUNE-SP, this gradient was predicted to be about 15 mK by the CFD simulations~\cite{pdsp_tdr,dune_tdr4}.

Even with homogeneous mixing, the CFD simulations predict that the concentration of impurities may vary across the cryostat volume, requiring a position-dependent correction to the electron lifetime. The Purity Monitors themselves are intrusive objects which cannot be deployed inside the active volume, but rather only at a few well-defined locations near the cryostat walls; thus, precise inference of the electron lifetime map requires alternative methods, as the one proposed in this article. A net of temperature sensors cross-calibrated to the $<$5 mK level should allow the measurement of a 15 mK temperature gradient, which can be used to constrain CFD simulations, providing a data-driven prediction of the impurity concentration. The main limitations of this new approach are the precision of the cross-calibration of the temperature sensors and the accuracy of the simulations. In this article, precise temperature monitoring for ProtoDUNE-SP will be described, with particular emphasis on sensor calibration in the laboratory.

\section{The temperature monitoring system of ProtoDUNE-SP}
\label{sec:protoDUNE}

\noindent ProtoDUNE-SP~\cite{pdsp_tdr} was the Single-Phase demonstrator of DUNE Far Detector HD module \cite{dune_tdr4}. 
The elements constituting the TPC, its associated readout electronics and the photon detection system, were housed in a 8x8x8 m$^3$ cryostat that contained the LAr that served both as target and detector material. The cryostat, a free-standing steel-framed vessel with an insulated double membrane, is based on the technology used for liquefied natural gas storage and transport. A cryogenic system maintains the LAr at a stable temperature of about 87 K by operating at a slight and constant over-pressure. It also ensures the required purity level by means of a closed-loop process that recovers the evaporated argon, recondenses it, filters it, and recirculates it back into the cryostat, keeping the LAr level at about 7.3 m from the bottom membrane. ProtoDUNE-SP was exposed to a charged particle test-beam from October to November 2018, and later recorded cosmic rays until January 2020~\cite{pdsp_1,pdsp_2}. It was finally emptied and decommissioned during Summer 2020.

\begin{figure}[htbp]
\begin{center}
\includegraphics[angle=-90, width=0.6\textwidth]{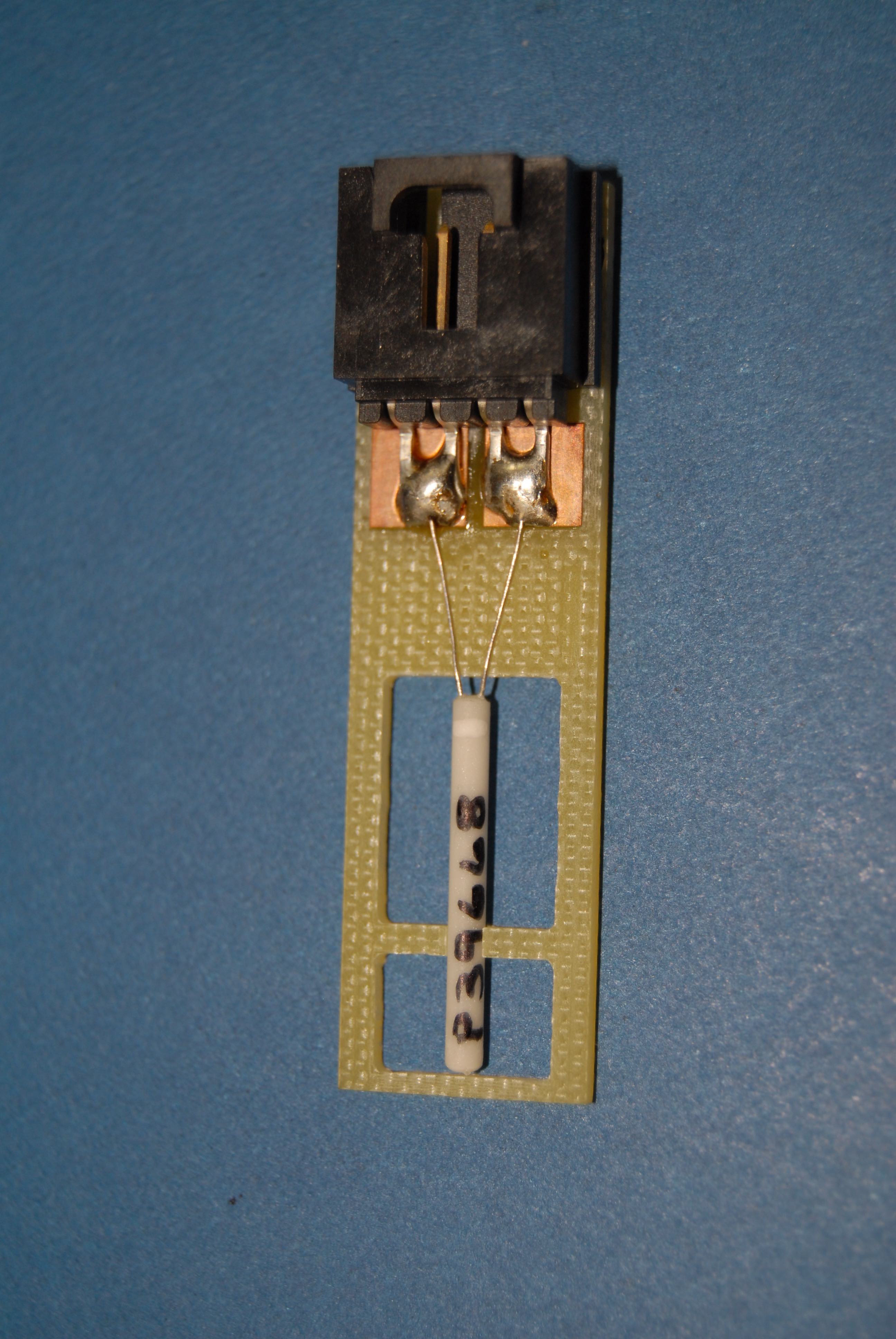}%
\caption{PCB support with temperature sensor and IDC-4 connector. The transition from two wires at the sensor to 4 wires at the readout is clearly seen. The sensor has a length of 2 cm.
\label{fig:sensor}}
\end{center}
\end{figure}

In order to understand the LAr behaviour inside the cryostat and validate the CFD simulations, 92 high-precision temperature sensors were installed inside ProtoDUNE-SP, near the active volume. These sensors were distributed in two vertical arrays, or Temperature Gradient Monitors (TGM), and two horizontal grids below and above the TPC, respectively. Three elements were common to all systems: sensors, cables and readout electronics. Resitance Temperature Detector (RTD) technology~\cite{minco} was chosen for this application. It consists of a metallic element whose resistance changes with temperature. This resistance is measured by feeding the RTD with a known current and measuring the resulting voltage. Based on previous experience from other prototypes~\cite{35t_1}, Lake Shore PT102 platinum sensors~\cite{pt102} with 100 $\Omega$ resistance at room temperature were chosen. Sensors were mounted on a 52x14 mm$^2$ PCB with an IDC-4 connector manufactured by Molex~\cite{bib:idc4male,bib:idc4female}, such that they could be plugged-in at any time. Several versions of the PCB have been explored, finally converging to the one shown in Fig.~\ref{fig:sensor}, which minimizes the contact of the sensor with the PCB while keeping the sensor protected.

A careful choice of the readout cable and the connections are essential to obtain the required temperature precision. See for example Ref.~\cite{minco} for a detailed description. A custom cable made by Axon~\cite{axon} was used. It consists of four American Wire Gauge (AWG) 28 teflon-jacketed copper wires, forming two twisted pairs, with a metallic external shield and an outer teflon jacket. The outer diameter of the cable is 3.7 mm. Teflon was chosen for its good thermal properties and low out-gassing. The metallic external shield (connected to the readout in one end and left floating at the sensor's end) and the twisted pairs are crucial to reduce the effect of external electromagnetic noise pick-up. When RTDs are far from the voltmeter, the resistance of cables and connectors are added to the one of the sensor, biasing the temperature measurement. This bias can be subtracted to some level, but cannot be fully controlled since the resistance of those elements also depends on temperature. To minimise the impact of this effect a four wire-terminal readout is employed~\cite{minco}, such that the voltage is measured in the vicinity of the RTD.

The last common element is the readout system, consisting of a very precise 1 mA current source to excite the sensors and a 24 bits ADC to measure the voltage. The readout system will be described in detail in Section \ref{sec:readout}.

As previously mentioned, ProtoDUNE-SP CFD simulations predict vertical temperature gradients as low as 15 mK~\cite{pdsp_tdr}. A relative precision better than 5 mK was required to validate and tune those simulations with sufficient confidence. Three main ingredients are necessary to obtain such a precision: i) cryogenically rated, high-accuracy temperature probes suitable for LAr applications., ii) a very precise readout system, and iii) accurate and stable long-term calibration, both for the readout and the sensors. As mentioned above, two TGMs were deployed in ProtoDUNE-SP: one could be moved vertically and hence calibrated \textit{in situ}, while the other one ---static--- fully relied on a prior calibration in the laboratory. In this article, the calibration of the static TGM~\cite{tfm} sensors is described in detail. This device consists of a vertical array of 48 sensors, installed 20 cm away from the lateral field cage.

The calibration was performed once in spring 2018  (a few months before the operation of ProtoDUNE-SP), and several times after ProtoDUNE-SP decommissioning. In this article the calibration setup, procedure and results will be described in detail. Comparison between the different calibration campaigns will be addressed, revealing important information about systematic uncertainties and RTD ageing.
\section{Readout system}
\label{sec:readout}
\begin{figure}[htbp]
\centering
\includegraphics[width=0.6\textwidth]{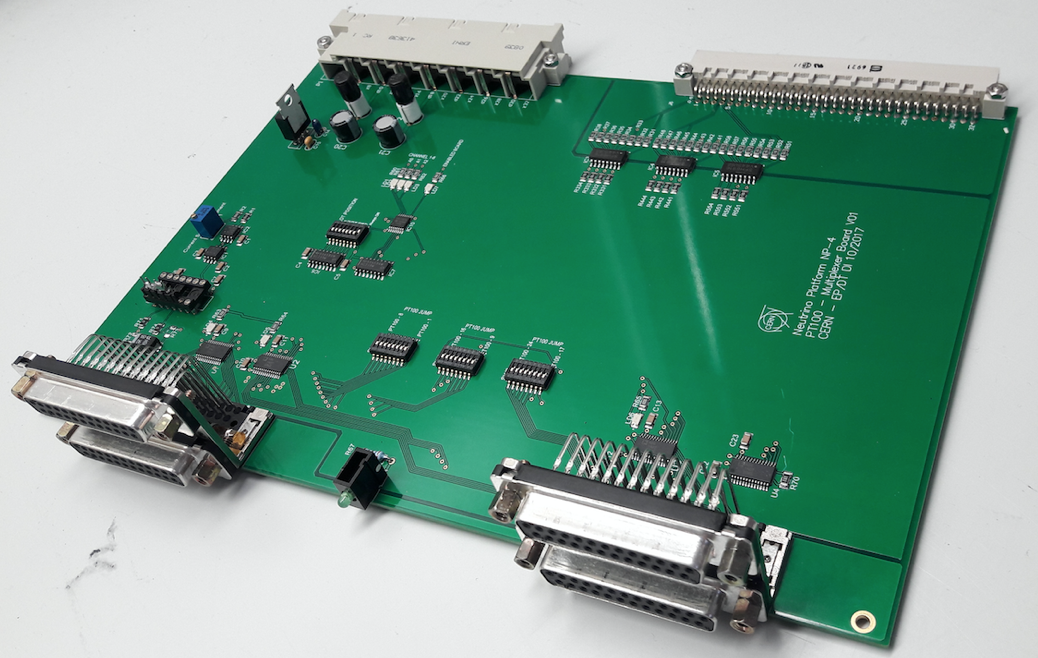}
\caption{Current source and multiplexing card with 24 channels.
\label{fig:readout}}
\end{figure}

\noindent A precise and stable electronic readout system is needed to achieve the required precision. In previous versions of the calibration system, each sensor was connected to a different and independent electronic circuit and thus, fed by a different current and read by a different ADC channel. It was soon realized that the measured temperature difference between any two pairs of sensors was heavily affected by the electronic offset between channels. This offset was not constant, and showed dependence on ambient temperature and humidity, which affected both the current source and the ADC, generating variations of tens of mK for the measurement of a single calibration constant between two sensors. A modified version of an existing PT100 mass temperature readout system, developed at CERN for one of the LHC experiments~\cite{bib:multiplexing_board}, was adapted to address this issue. The system consists of an electronic circuit that includes:

\begin{itemize}
\item A precise and accurate 1 mA current source for the excitation of the temperature sensors based on an application of the Texas Instruments precise voltage reference REF102CU with a possibility to adjust $\pm$10nA with Keithley 2001 multimeter~\cite{xavier,keithley}.
\item A multiplexing circuit based on the Analog Devices ADG1407BRUZ multiplexer with ultralow internal resistance in an 8-channel differential configuration. By integrating three multiplexers, the readout circuit supports simultaneous acquisition from 24 channels. The multiplexing circuit and the current source are assembled on a single card (see Fig.~\ref{fig:readout}).
\item A readout system based on National Instruments Compact RIO-FPGA device~\cite{compactrio} equipped with a NI-9238 analogue input module that provide 24 bits resolution over 1 volt range~\cite{ni9238}. By programming the Van Dusen equation the readout calculates the temperature in Kelvin units. The Compact RIO also drives the control bits of the multiplexers
\end{itemize}

\noindent
One of the features of the readout circuit is the serialization of the current source excitation for all the sensors connected to the same board, such that the same current is delivered to all of them. Multiplexing the signal of the sensors such that they can be readout by the same ADC channel minimises the residual offset due to the electronics. This system was used during the calibration campaigns presented in this article and also for the temperature measurements during ProtoDUNE-SP operation.

\begin{figure}[htbp]
\centering
{\includegraphics[width=0.65\textwidth]{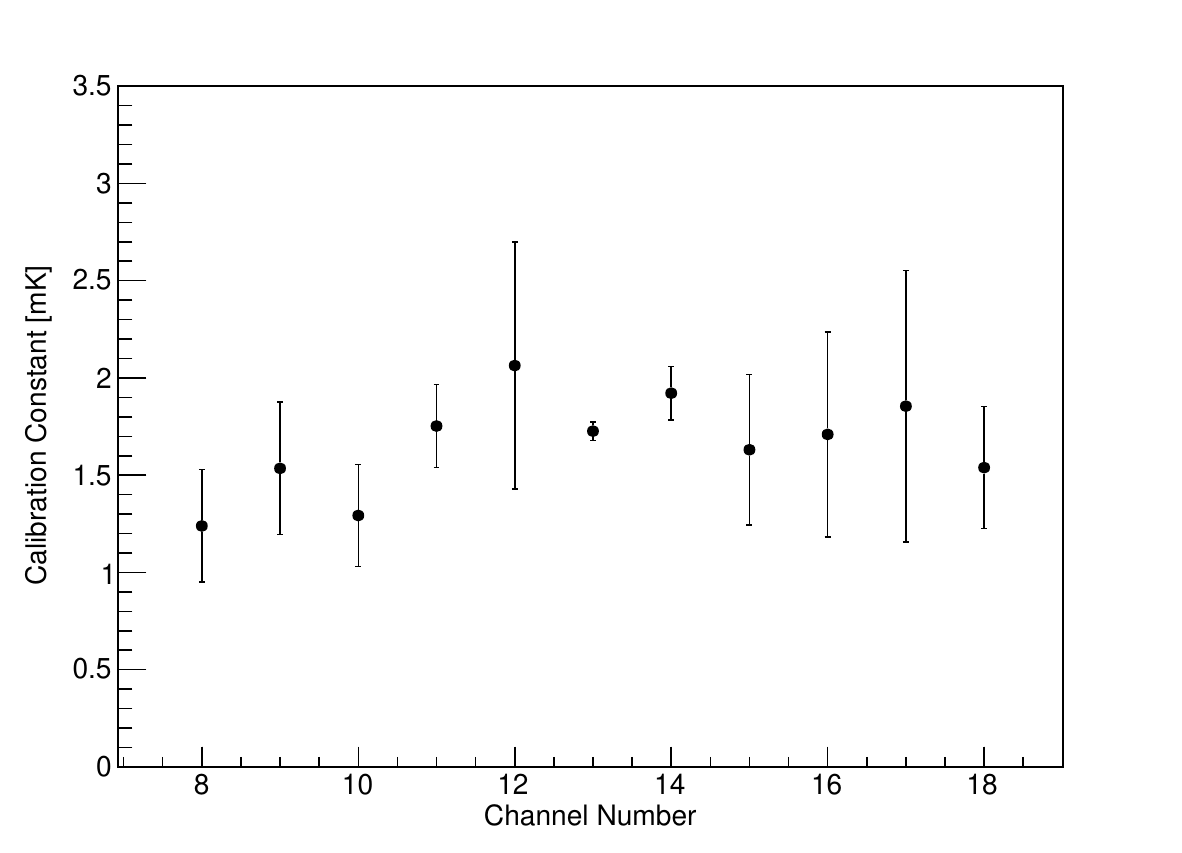}}
\caption{Offset between readout channels 8-18 and channel 7, used as reference. Points correspond to the mean of the 4 independent measurements and the error bars are their standard deviation.}
\label{fig:readout_calib}
\end{figure}

The readout was not considered a potential source of bias during the calibration campaign prior to the installation in ProtoDUNE-SP (see Sec.~\ref{sec:old_calib}) and it was not studied in detail at that time. However, in order to improve the calibration results following ProtoDUNE-SP operation, a more detailed study of the readout system was conducted prior to the calibration campaigns presented in~\ref{sec:new_calib}. In particular, the twelve channels (7-18) used for sensor calibration were investigated. It was found that, despite the use of the multiplexing system, a small residual offset between channels persists. Fig.~\ref{fig:readout_calib} shows the offsets of channels 8-18 with respect to channel 7, computed using twelve 20-ohm precision resistors with a low Temperature Coefficient of Resistance (TCR)\footnote{The temperature coefficient of resistance is defined as the change in resistance per unit resistance per degree rise in temperature. Typically ±5ppm/°C.} precision resistors, with an equivalent temperature of 76 K. Two of those, High Precision 1 (HP1) and High Precision 2 (HP2), are selected as the measurement samples and connected to the reference channel (7) and the channel being calibrated (from 8-18), while leaving all other channels connected to other secondary resistors to let the current flow through the system. The readout offset between channel 7 and channel X, $\Delta T_{7-X}^{readout}$, is then computed using the results of two consecutive measurements:

\begin{enumerate}
    \item HP1 in channel 7 and HP2 in the channel X being calibrated. The offset between the measured temperatures is

    \begin{equation}
        \Delta T_{7-X}^A = T_{X}^{HP2}-T_{7}^{HP1}+\Delta T_{7-X}^{readout} \, .
    \end{equation}

    \item HP2 channel 7 and HP1 in the channel X being calibrated. The offset between the measured temperatures is
\end{enumerate}

    \begin{equation}
        \Delta T_{X-7}^B = T_{X}^{HP1}-T_{7}^{HP2}+\Delta T_{7-X}^{readout} \, .
    \end{equation}

\noindent
Because of the very low TCR of the resistors, it can be assumed that the resistances are constants in thse two measurements ( $T_{7}^{HP1}=T_{X}^{HP1}$ and $T_{7}^{HP2}=T_{X}^{HP2}$). Thus, the offset between channels 7 and X can be computed as the average between those two measurements:

\begin{equation}
    \Delta T_{7-X}^{readout} = \frac{\Delta T^{A}_{7-X} + \Delta T^{B}_{X-7}}{2} \, .
\end{equation}

The results in Fig.~\ref{fig:readout_calib}, show an offset of up to 2.5 mK when comparing directly with channel 7, while the offset between any other two channels is below 1 mK, indicating a special feature of this channel. Error bars in that figure correspond to the standard deviation of four independent measurements (repeatability) of the same offset (see Fig.~\ref{fig:readout_calib}). As it can be observed the error is below 0.5 mK, what probes the great repeatability of the readout. These offsets are more likely due to parasitic resistances in the different lines that are multiplexed. This finding allowed a correction to the measurements taken during the subsequent calibration campaigns, improving the obtained precision.

\section{Calibration before the installation in ProtoDUNE-SP}
\label{sec:old_calib}

\noindent The Lake Shore Cryotronics company provides PT102 RTDs with a temperature accuracy of about 0.1 K, which is insufficient for ProtoDUNE-SP's requirements. While the company offers additional calibration to the 10 mK level, the cost is prohibitive. R\&D on sensor calibration was identified as a crucial ingredient for the success of the ProtoDUNE-SP temperature monitoring system. For this particular application, sensor calibration consists in finding the temperature offset between any pair of sensors when exposing them to the same temperature. The experimental setup used for the calibration of RTDs before their installation in ProtoDUNE-SP had two main components: i) the readout system used to monitor the sensor's temperature (described in the previous section) and ii) the cryogenics vessel and the associated mechanical elements used to put sensors together under stable and homogeneous cryogenics conditions. They were developed to achieve a relative calibration with a precision better than 5 mK. 

\subsection{Experimental setup}
\begin{figure}[htbp]
\centering
\includegraphics[width=0.32\textwidth]{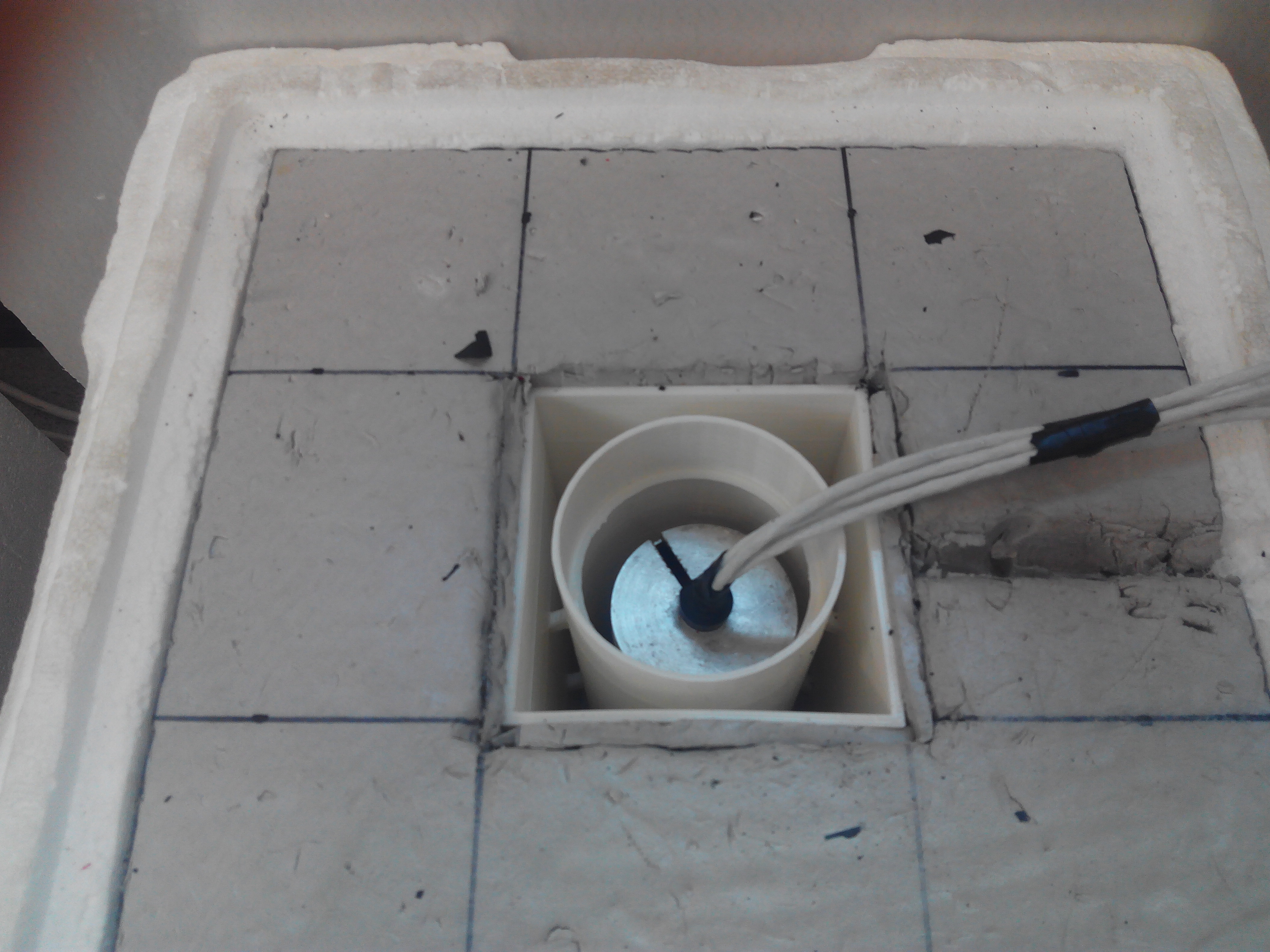}
\includegraphics[width=0.32\textwidth]{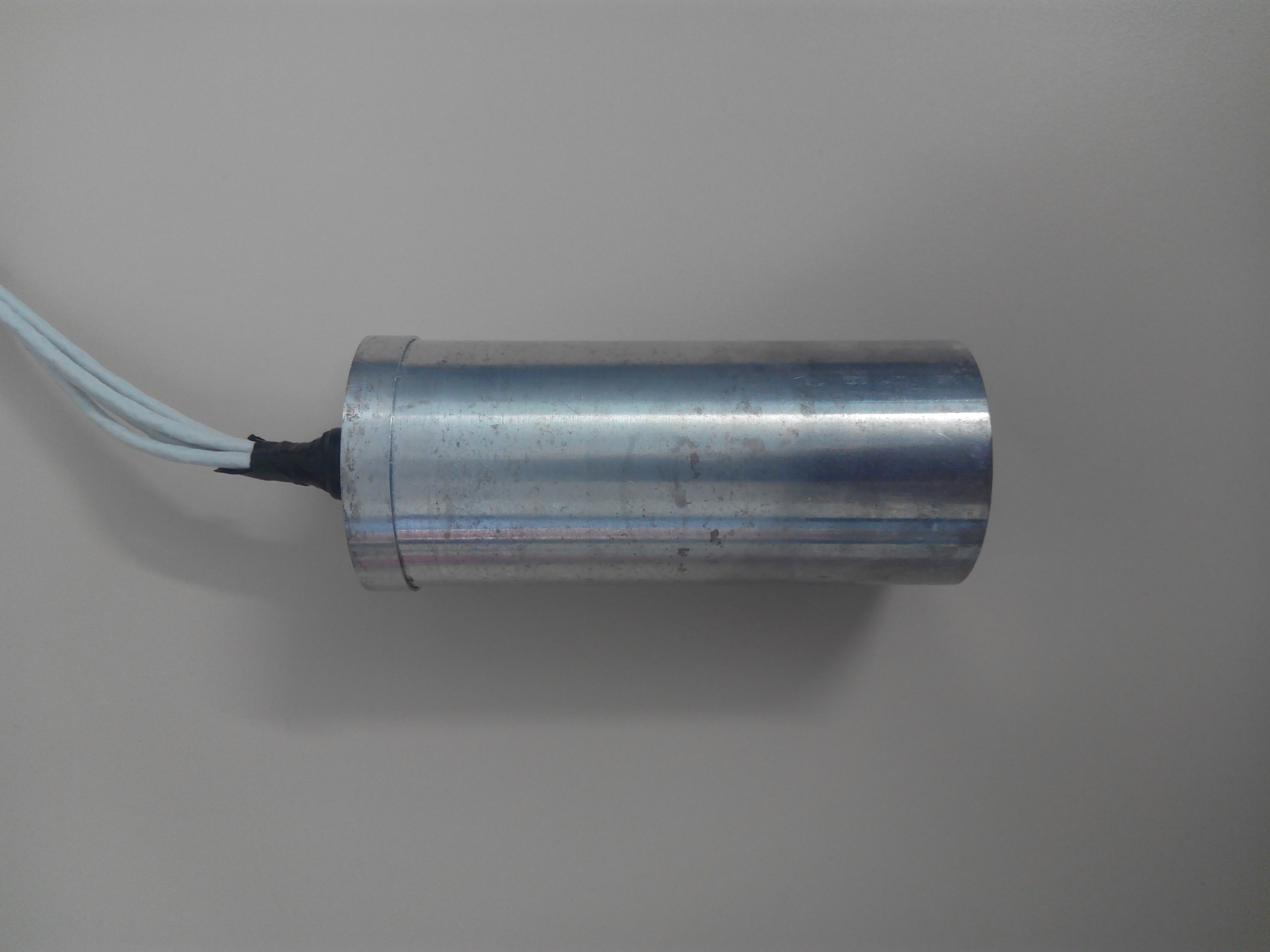}
\includegraphics[width=0.32\textwidth]{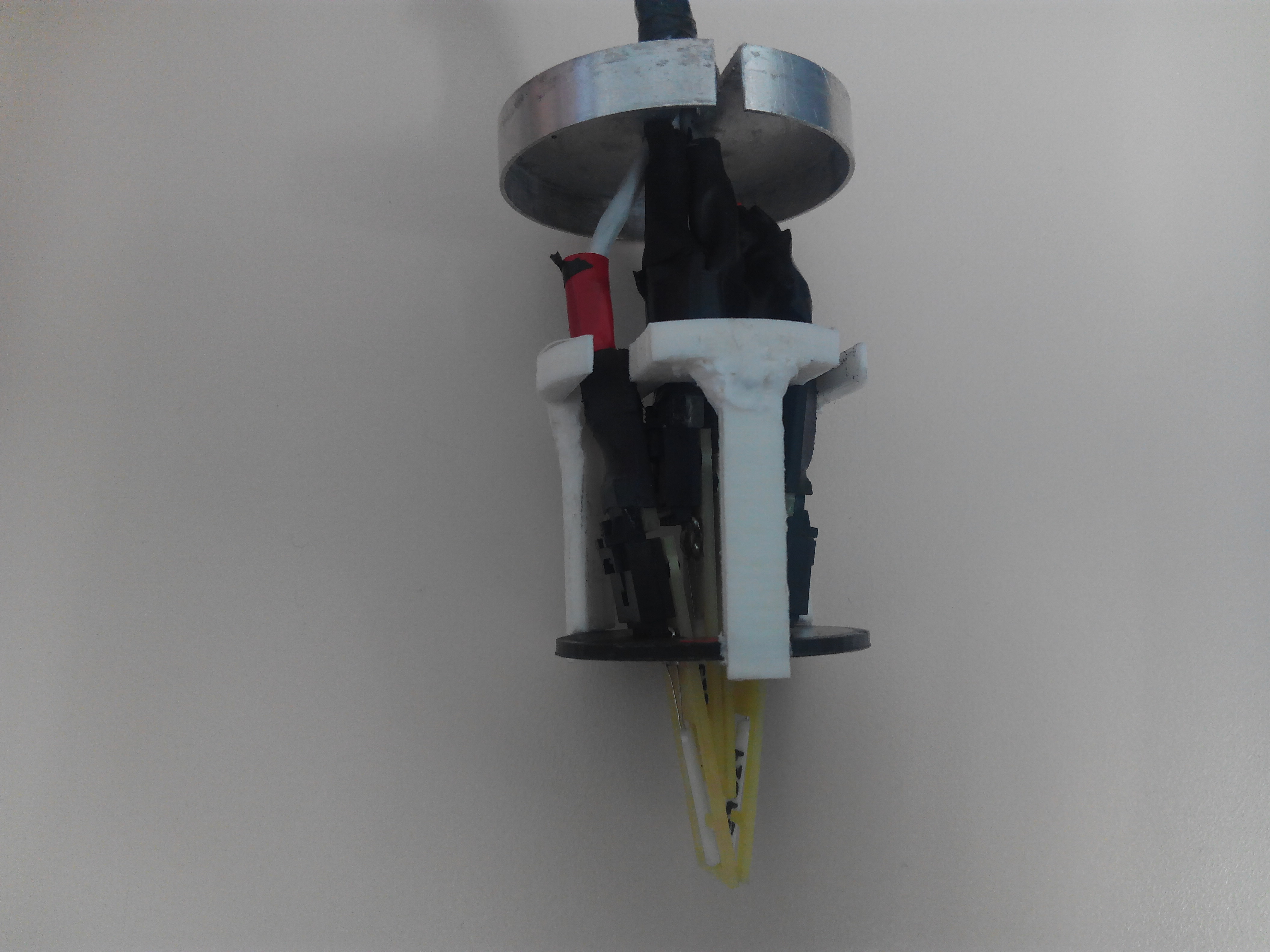}
\caption{Final calibration setup. Left: polystyrene box with PLA box and aluminum capsule. Middle: aluminum capsule. Right: Sensor's support with four sensors.
\label{fig:setup_final}}
\end{figure}

\noindent The mechanics of the calibration setup evolved significantly from the initial tests to the final Static TGM  calibration campaign. This evolution was primarily driven by the need to improve offset stability and repeatability. The final configuration (see Fig. \ref{fig:setup_final}) consisted of the following components:

\begin{itemize}
    \item A polystyrene vessel formed by an outer box with dimensions $35\times35\times30$ cm$^3$ and 4.5 cm thick walls, and a dedicated polystyrene cover, complemented by extruded polystyrene panels glued into the inner walls and floor of the outer box, to conform an inner empty volume of $10\times10\times20$ cm$^3$.
    \item A $10\times10\times20$ cm$^3$ 3D printed polylatic acid (PLA) box with two independent concentric volumes, placed in the inner volume of the polystyrene vessel. Its purpose is twofold: i) to contain LAr, since polystyrene is porous to it, and ii) to create an smaller inner volume  with further insulation and less convection.
    \item A cylindrical aluminum capsule, to be placed in the inner volume of the PLA box,  with 5 cm diameter, 12 cm height and 1 mm thin walls. It had a circular aluminum cover with a small opening to extract the cables and to allow LAr to penetrate inside. The capsule was used to slowly bring sensors to cryogenic temperatures by partial immersion in LAr with no liquid inside, minimizing thermal stress. Aluminum was chosen for its high thermal conductivity.
    \item  A 3D printed PLA support for four sensors, to be placed inside the aluminum capsule, keeping sensors always in the same position with respect to each other and to the capsule walls.
\end{itemize}

\begin{figure}[htbp]
\centering
\includegraphics[width=0.9\textwidth]{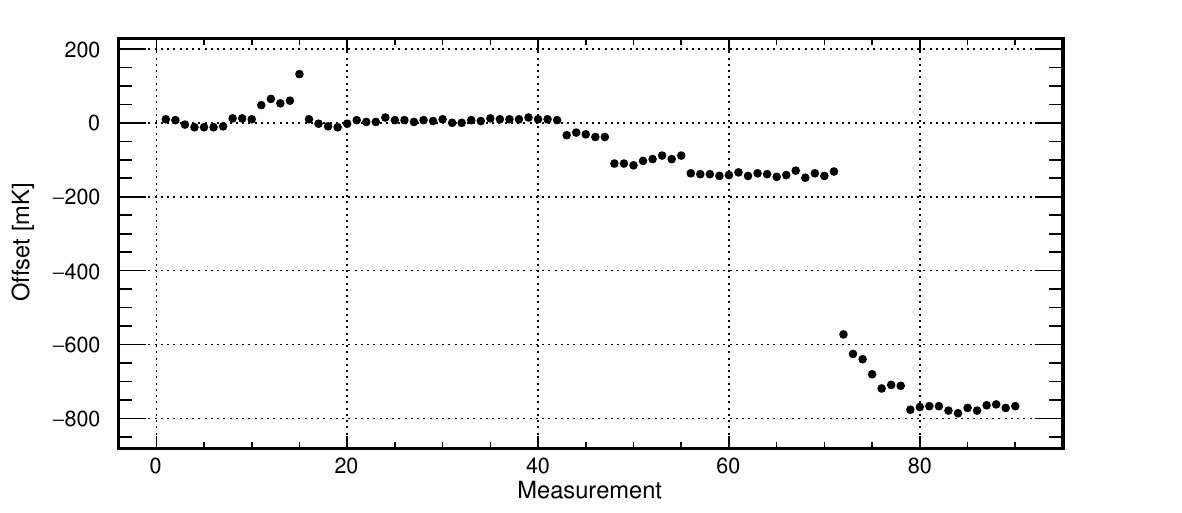}
\caption{Offset between two sensors for 90 immersions in LN2. A temperature drop of approximately 800 mK is observed around the 70th immersion point.
\label{fig:broken_sensor_evolution}}
\end{figure}

\begin{figure}[htbp]
\centering
\includegraphics[width=0.6\textwidth]{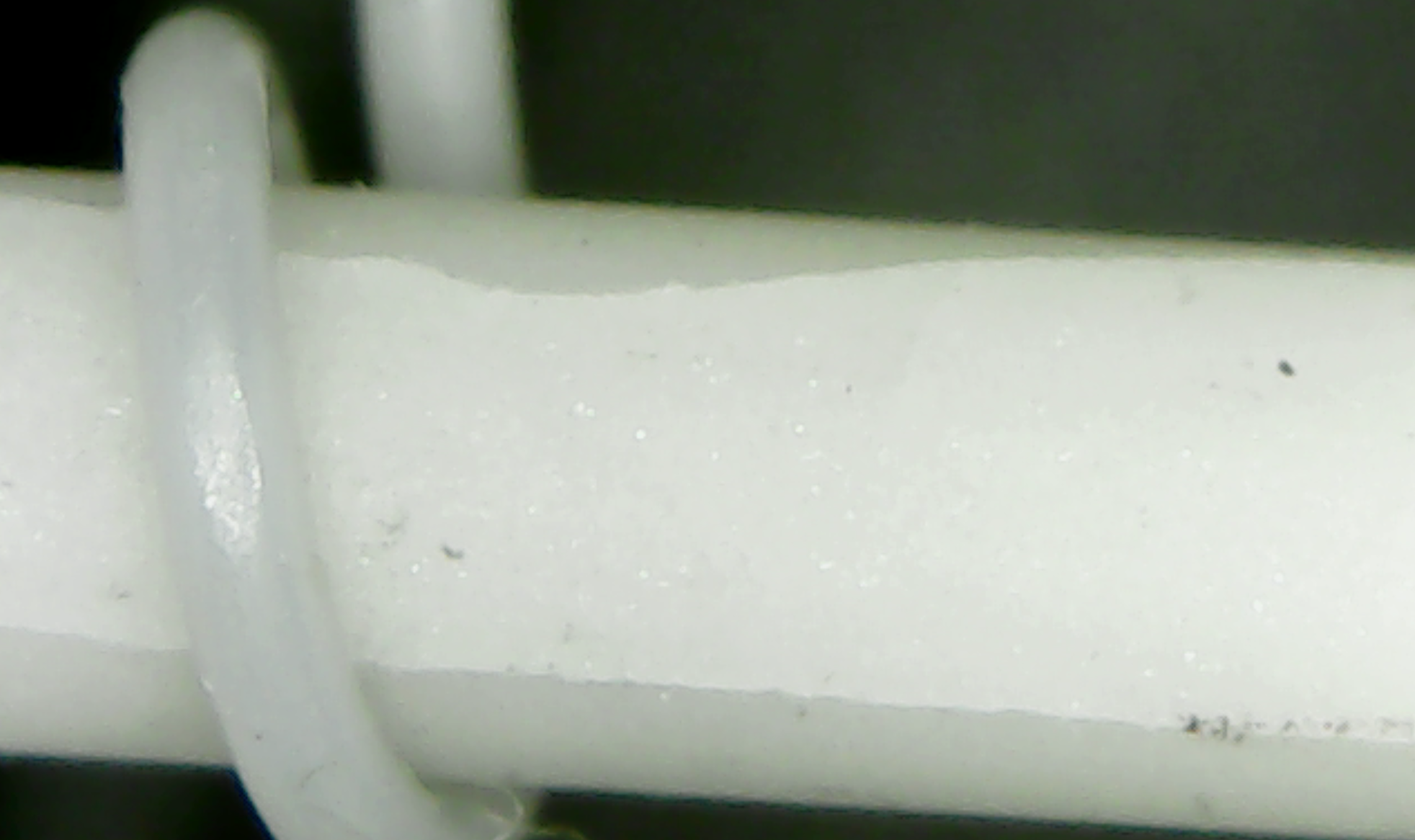}
\caption{Cracks observed with the microscope in the ceramic of the sensor.
\label{fig:broken_sensor}}
\end{figure}

The system described above ensures sufficiently stable and homogeneous conditions within the inner volume, with three levels of insulation: the outer polystyrene vessel and two PLA box LAr volumes. The aluminium capsule is key to this system and its usage constituted a turning point in the R\&D since it minimises thermal shocks, which were identified as the main limiting factor for the repeatability of the sensor's offsets. Indeed, variations of several tenths of mK were observed during initial tests without the capsule. The problem was attributed to thermal shocks when, after many immersions in LN$_{2}$, one of the sensors suffered a dramatic change in its offset (see Fig.~\ref{fig:broken_sensor_evolution}). Examination at the microscope revealed cracks in the outer RTD ceramics (see Fig.~\ref{fig:broken_sensor}).

\subsection{Calibration procedure}
\label{sec:calib_procedure}

\begin{figure}[htbp]
\centering
\includegraphics[width=\textwidth]{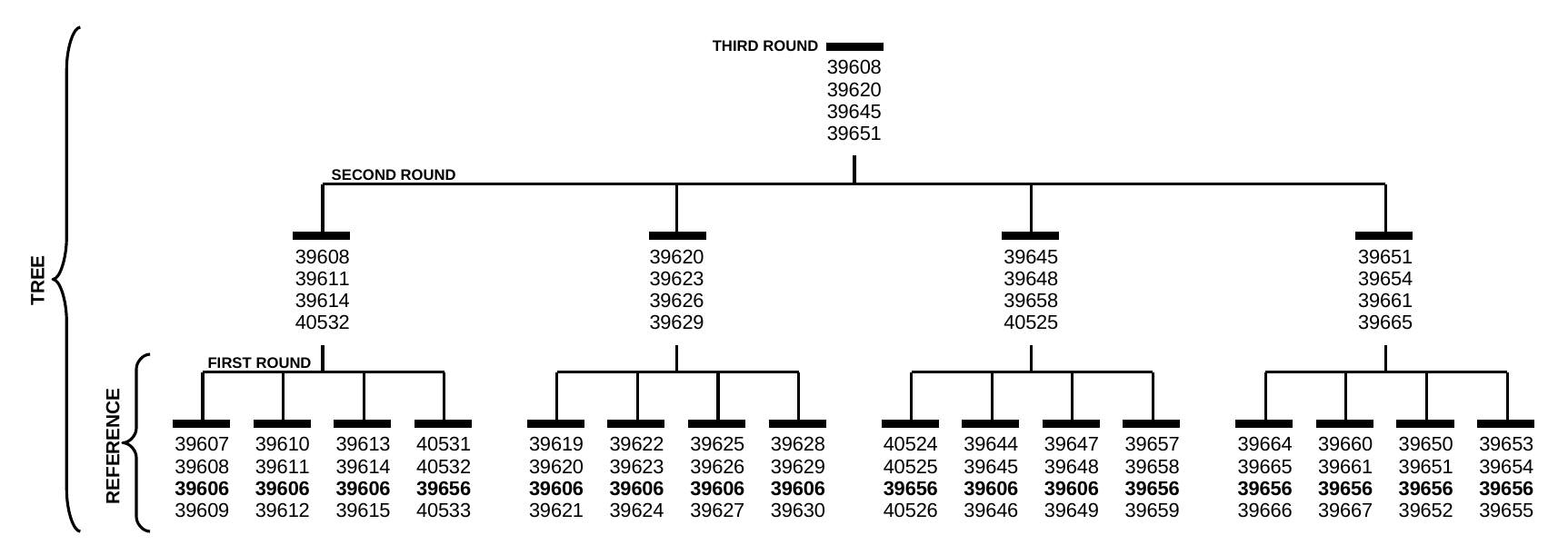}
\caption{Schematic view of calibration sequence. Each number represents a different RTD, and the numbers in bold represent the reference sensors. The first round consists on a direct calibration respect to a reference sensor. The second and third rounds are needed to relate all sensors independently of the reference sensor.
\label{fi:CAL_sequence}}
\end{figure}

\noindent The calibration procedure relies on the assumption that all sensors in the capsule are at the same temperature. This limits the number of RTDs in a single run to four, since i) they should be as close as possible to each other and ii) as far as possible from the capsule walls (the sensor closer to the wall could be biased). Two different methods, described schematically in Fig.~\ref{fi:CAL_sequence}, were used to cross-calibrate all 48 sensors in runs of 4 sensors:

\begin{enumerate}
\item {\bf Reference method}: all sensors are calibrated with respect to a reference one, in sets of three sensors (the fourth one would be the reference sensor, which must be present in all runs). In total there are sixteen calibration sets.
\item {\bf Tree method}: Four different sets of sensors can be cross-calibrated by performing a second round of measurements with a single promoted sensor from each of those four sets. Since there are 16 sets in total, a third round of measurement is needed to cross-calibrate the four sets in the second round.
\end{enumerate}

The reference method was expected to be more precise since any two sensors were related through a single intermediate sensor (the reference one), while for the tree method the relation between any two sensors required more than one intermediate sensor. For example, the offset between sensors $i$ and $j$ in different sets, related through sensors $k$ and $l$ in the second round, would be $\Delta T_{ij} = \Delta T_{ik} + \Delta T_{kl} + \Delta T_{lj}$. For sensors related through the third round an additional term should be added, further increasing the offset uncertainty.

Offset repeatability is a critical parameter to understand, as the laboratory calibration must remain valid when applied to the actual detector. To study this, four independent calibration runs were performed for each set of four sensors. However, due to concerns about thermal fatigue in the primary reference sensor---subjected to 64 thermal cycles over the four repeatability runs---three secondary reference sensors were employed to regularly monitor its response and ensure reliability over time.

\begin{figure}[htbp]
\centering
\begin{tabular}{ c }
\includegraphics[height=0.34\textheight]{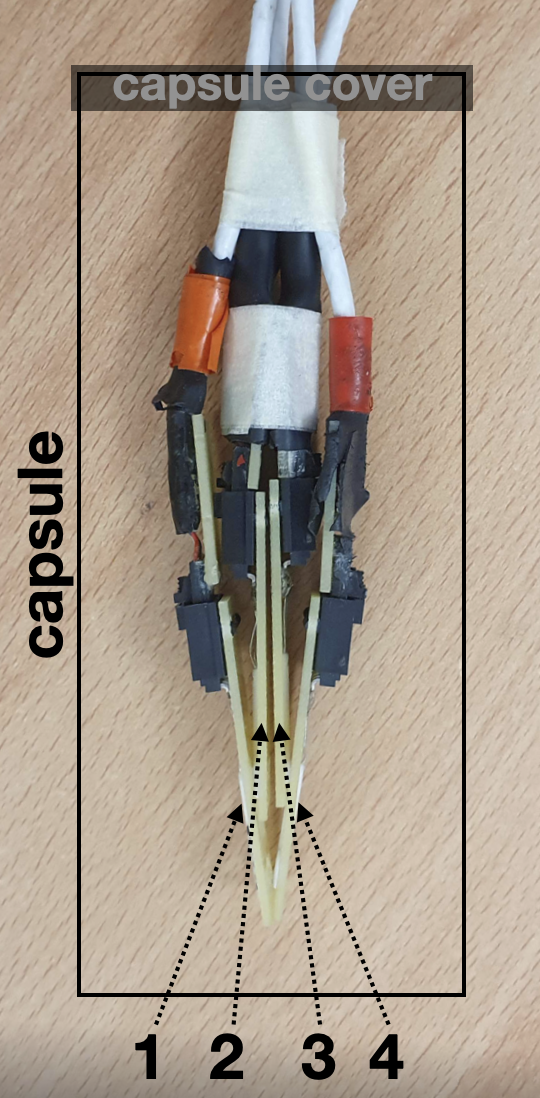}
\end{tabular}%
\begin{tabular}{ c }
\includegraphics[height=0.1665\textheight]{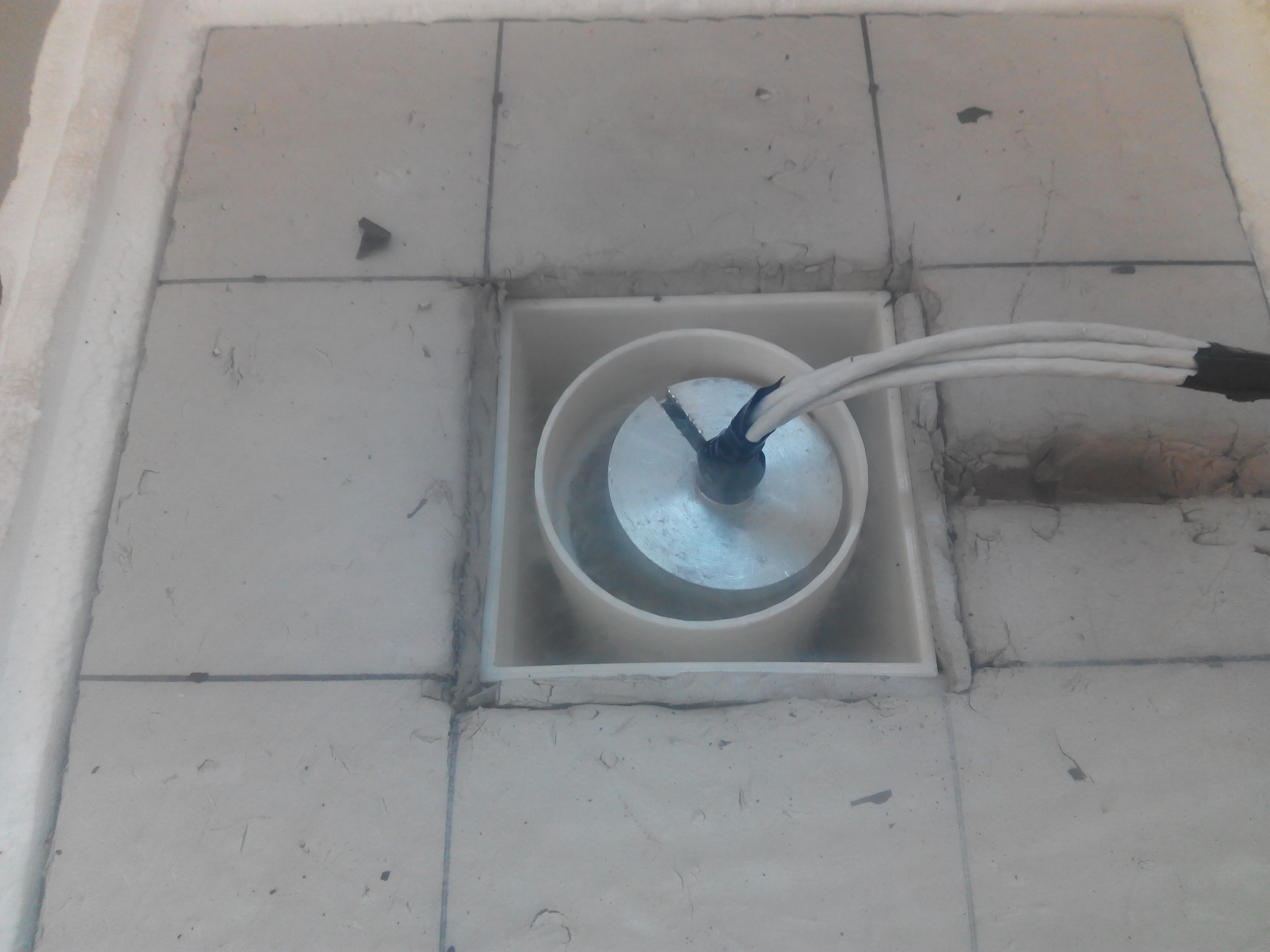} \\
\includegraphics[height=0.1665\textheight]{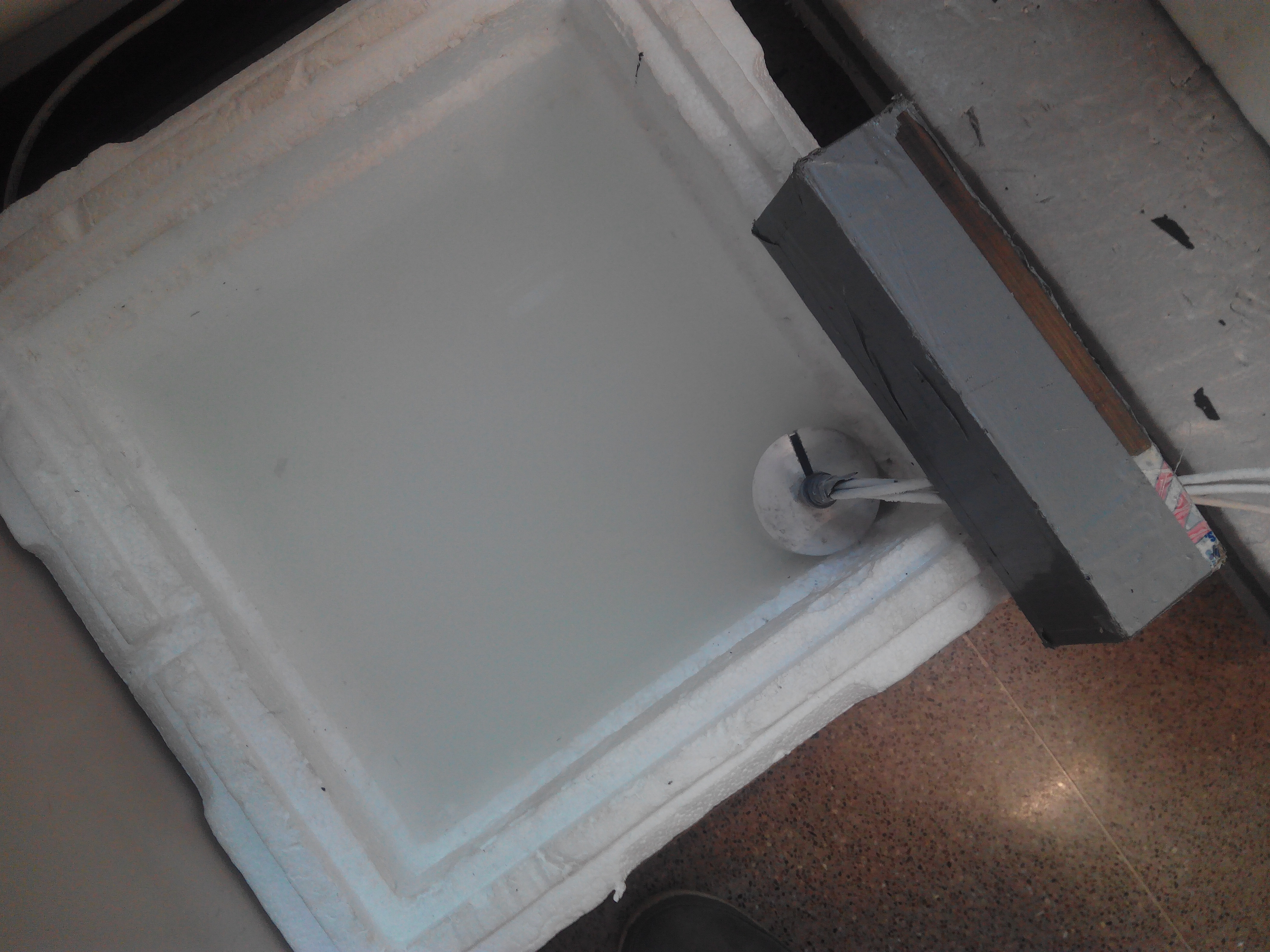}
\end{tabular}
\caption{Left: Picture of sensor order inside the flask. Position 1 corresponds to highest serial number and 4 to lowest; position 3 is reserved to reference sensor.
 Top: Flask partially introduced in LAr.
 Bottom: Flask partially introduced in room water to warm up sensors before performing a new measurement.
\label{fi:CAL_procedure}}
\end{figure}

The following procedure is applied for each set of four sensors. First, they are placed inside the aluminium capsule. The central connector closer to the red one is used for the reference sensor. These positions are easily called by numbers from 1 (orange) to 4 (red), as shown in Fig.~\ref{fi:CAL_procedure}-left.

After filling the PLA box with liquid argon, the aluminum capsule is partially immersed (see Fig.~\ref{fi:CAL_procedure}-right-top), allowing the air inside to cool down gradually. This setup prevents the sensors from coming into direct contact with liquid argon, instead inducing a slow cooldown through the surrounding cold gas atmosphere, thereby avoiding thermal shock. When the monitored temperature approaches the one of LAr ($< 95$ K), which takes about 15 minutes, the capsule is completely immersed, being filled with liquid. The PLA box is then completely filled with LAr to account for evaporation during the cool-down phase. Finally, the polystyrene vessel is closed using the polystyrene lid. The calibration run actually begins at that point and lasts for 40 minutes.

When the measurement is finished, the capsule is extracted from the liquid, emptied of LAr and partially immersed into another polystyrene box filled with water at room temperature (see Fig.~\ref{fi:CAL_procedure}-right-bottom). The warm-up process takes about ten minutes. When sensors have reached a temperature around 250 K a new independent calibration run can start.

\begin{figure}[htbp]
\centering
{\includegraphics[width=0.6\textwidth]{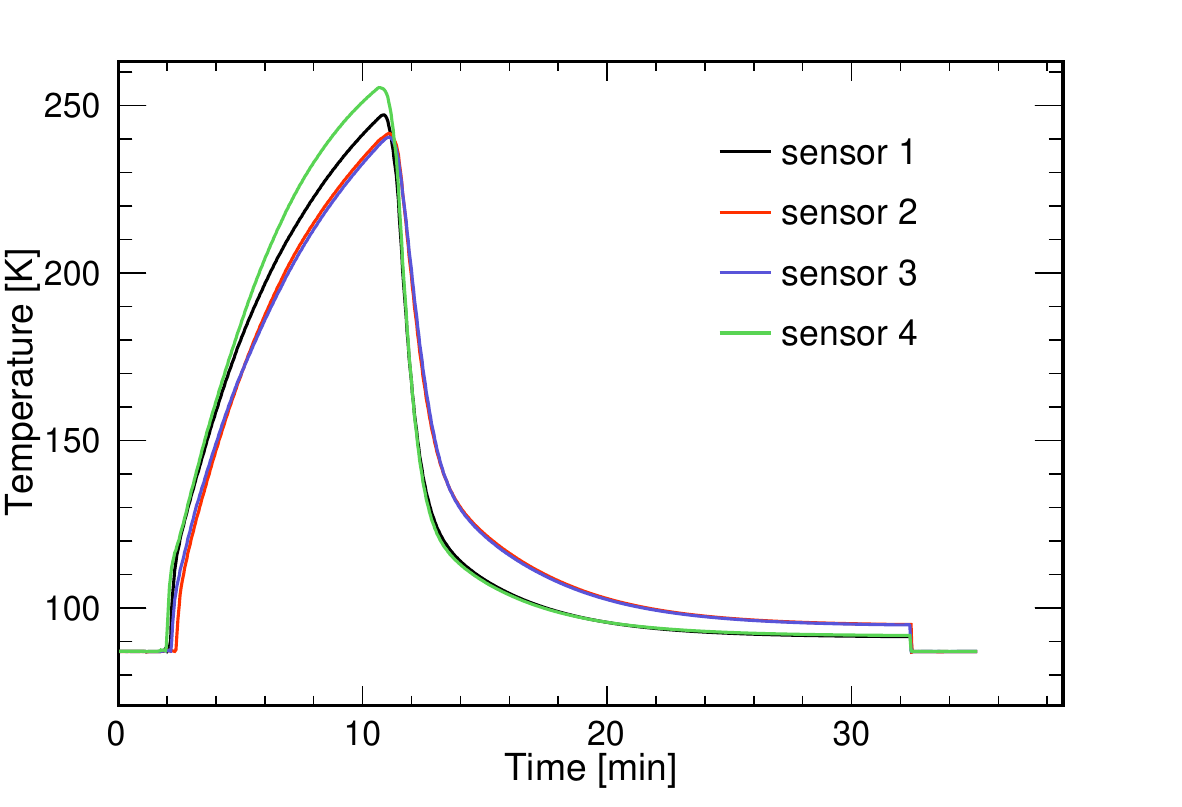}}
\caption{Temperature evolution between two calibration runs, showing the warm-up and cool-down phases of the four temperature sensors inside the capsule. Sensors 1 and 4 are located closest to the capsule walls.}
\label{fi:CAL_pre}
\end{figure}

Fig.~\ref{fi:CAL_pre} shows the typical evolution of the temperature for the four sensors in the capsule during the warm-up and cool-down processes. A sudden fall to 87 K is observed at minute 32, which corresponds to total immersion of the sensors in LAr. Notice also that two of the sensors have lower temperature during the cool-down process; those are sensors 1 and 4, the ones closer to the capsule walls.

For each set of sensors the procedure described above is repeated four times, resulting in four independent measurements of the same offset. These measurements are used to compute a mean offset value and its standard deviation (repeatability from now on).

\subsection{Calibration results}
\label{sec:calib_results}
\noindent Here we present the results of the calibration using both calibration methods, a study of the consistency of these results, and a preliminary estimation of the systematic uncertainty of the calibration process.

\subsubsection{Results on the first round of measurements}
\begin{figure}[htbp]
\centering
{\includegraphics[width=\textwidth]{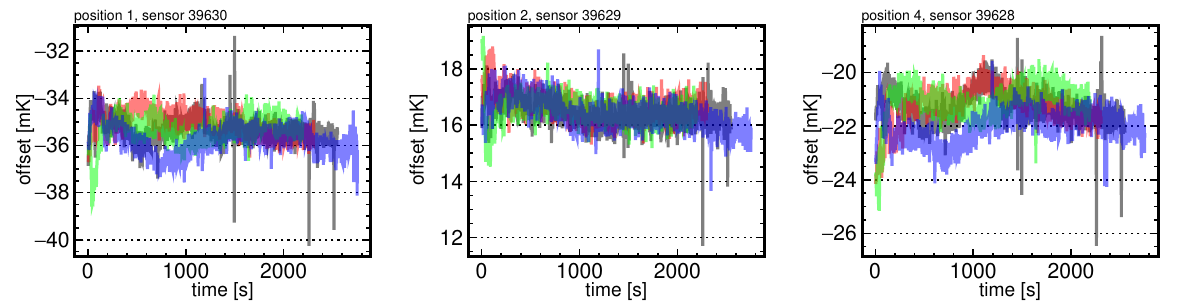}}
\caption{Offset between the reference sensor and each of the three sensors in a set, as a function of time. Each colour represents an independent calibration run.}
\label{fi:CAL_offset_example}
\end{figure}

\noindent Fig.~\ref{fi:CAL_offset_example} shows the offset of three different sensors with respect to the reference sensor as a function of time, and for four independent calibration runs. The offset is more stable for the sensor closer to the reference (position 2), while external sensors (positions 1 and 4) present larger variations, but show similar patterns between them. This effect is attributed to the geometry of the system, with sensors at different heights and not symmetrically positioned with respect to the capsule walls. This was taken into account when developing the second version of the system, presented in Sec.~\ref{sec:new_calib}.

\label{sec:results_first_round}

\begin{figure}[htbp]
\centering
{\includegraphics[width=0.6\textwidth]{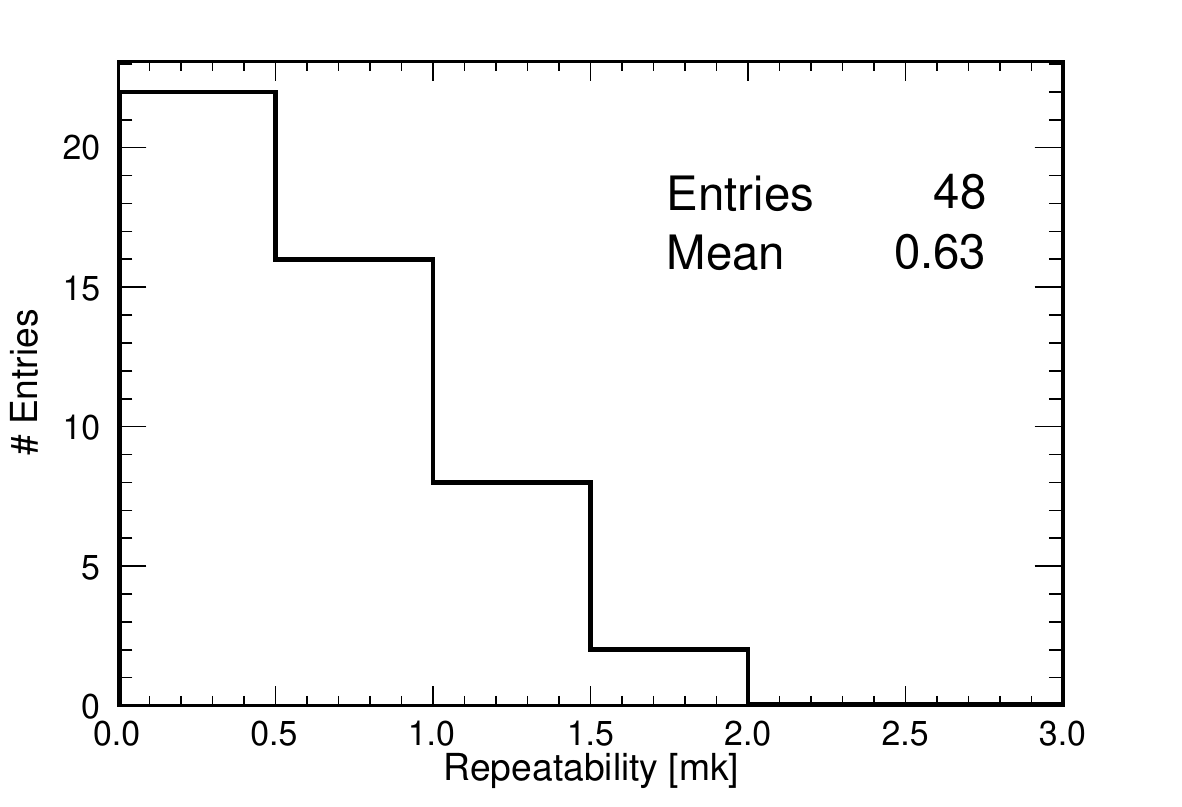}}
\caption{Distribution of the repeatability for calibration runs in the first round.}
\label{fi:CAL_rms_1r}
\end{figure}

The mean offset for each sensor and each calibration run is calculated as the average over the time interval between 1000 and 2000 seconds, identified as the most stable region for the majority of the runs. The standard deviation of the four means is taken as the uncertainty, hereafter referred to as repeatability. As shown in Fig.~\ref{fi:CAL_rms_1r}, the uncertainties are generally below 1~mK, demonstrating the high level of repeatability achieved in the calibration process.

\subsubsection{Time walk correction for reference sensors}
\label{sec:reference_corrections}
\begin{figure}[htbp]
\centering
{\includegraphics[width=0.48\textwidth]{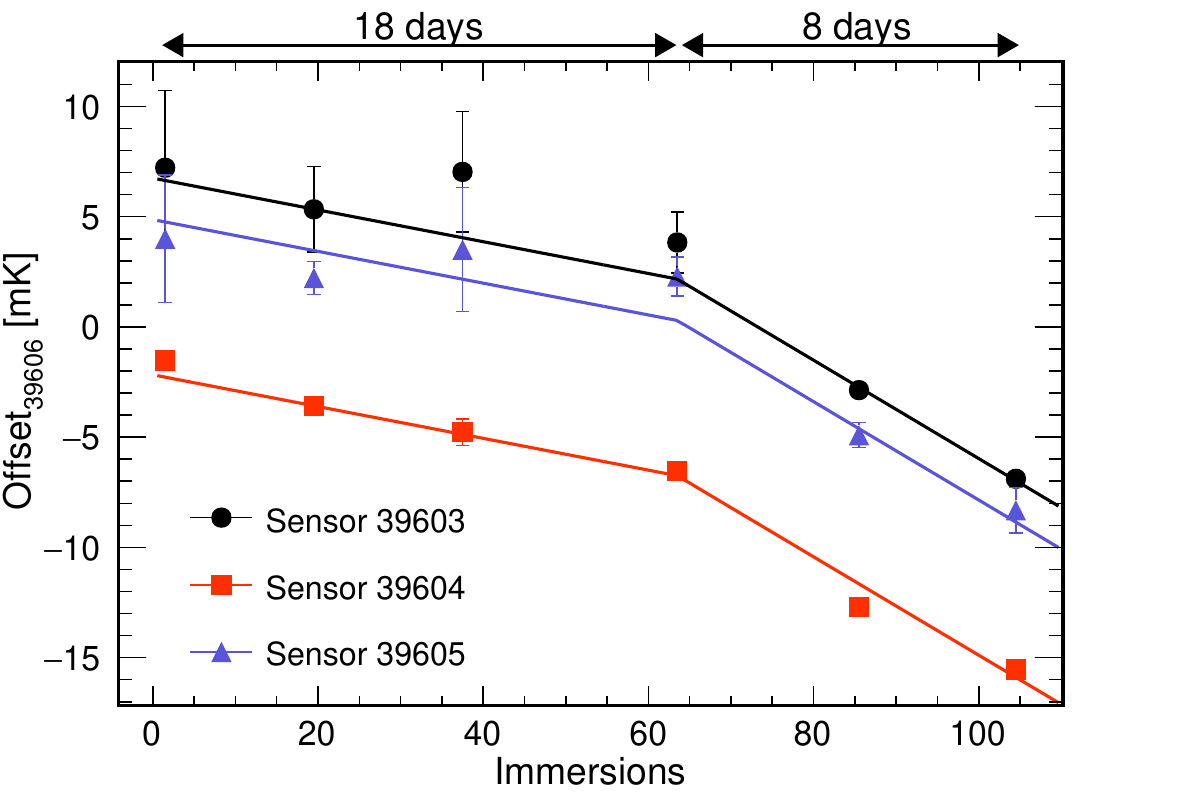}}
{\includegraphics[width=0.48\textwidth]{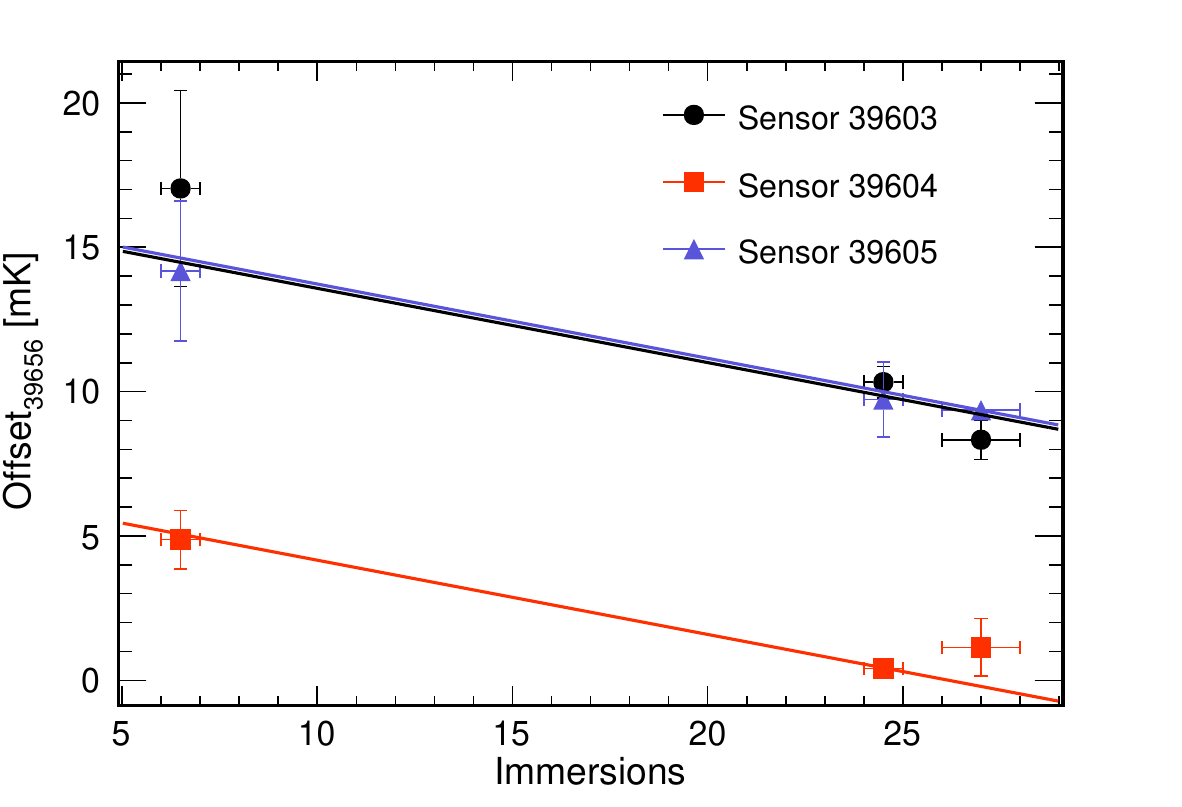}}
\caption{Offset between the reference sensors (39606 on the left panel and 39656 on the right panel) and the three secondary references as a function of the number of immersions. Offsets for sensor 39604 have smaller errors because of its position inside the capsule. Solid lines correspond to the parametrization in Eq.~\ref{eq:ref1_param}.}
\label{fig:offset_ref}
\end{figure}

\noindent Time evolution of the response of sensor 39606 was studied by periodically (every $\sim$20 immersions)  computing its offset with respect to three secondary reference sensors, 39603, 39604 and 39605. For each of those additional calibrations, two runs were taken instead of four in order to minimize thermal fatigue of the reference sensor. As shown in Fig.~\ref{fig:offset_ref}-left, the offset, computed as the mean of those two runs, varies linearly at a rate of 0.07 mK/immersion, which suddenly increases to 0.22 mK/immersion after 60 runs. This change in the slope may be related with to the frequency of immersions, which increased from 3/day to 5/day. Sensor 39606 was initially used for other purposes, and the calibration of the 48 sensors conforming the static TGM started approximately at immersion 40. Given this change in its response, sensor 39606 was substituted by 39656 as primary reference for the last quarter of the TGM calibration in order to avoid further fatigue and potential untraceable behaviour. The evolution of the new reference sensor is shown in Fig.~\ref{fig:offset_ref}-right. It is worth noting that the same slopes are valid for the three secondary reference sensors in both cases, supporting the idea that the change observed in the offsets can be exclusively attributed to thermal fatigue of the reference sensor.

The zero intercept in both panels of Fig.~\ref{fig:offset_ref} corresponds to the unbiased offset (or the offset at $N_{inmersions}=0$) of each of the secondary reference sensors with respect to the primary reference. In order to compute the unbiased offset for a sensor $s$, a time walk correction after $N$ immersions is parameterised as:

\begin{equation}
\Delta T_{s,06}(N=0)=
    \begin{cases}
        \Delta T_{s,06}(N)+(0.072\pm0.003)*N                             & N<63.5\\
        \Delta T_{s,06}(N)+(0.072\pm0.003)*63.5+\\+(0.223\pm0.007)*(N-63.5) & N>63.5 \,,
    \end{cases}
    \label{eq:ref1_param}
\end{equation}

\begin{equation}
\Delta T_{s,56}(N=0)=\Delta T_{s,56}(N)+(0.168\pm0.007)*N  \, .
\label{eq:ref2_param}
\end{equation}

All sensors can be related to the same primary reference sensor, 39606, adding  the unbiased offset between sensors 39606 and 39656,  $\Delta T_{56,06}(N=0)$, to sensors calibrated with respect to 39656. By averaging over the three secondary reference sensors, the value obtained is $\Delta T_{56,06}(N=0)=-9.19\pm0.13$ mk.

\subsubsection{Results on the reference method}
\label{sec:results_reference}
\begin{figure}[htbp]
\centering
{\includegraphics[width=0.9\textwidth]{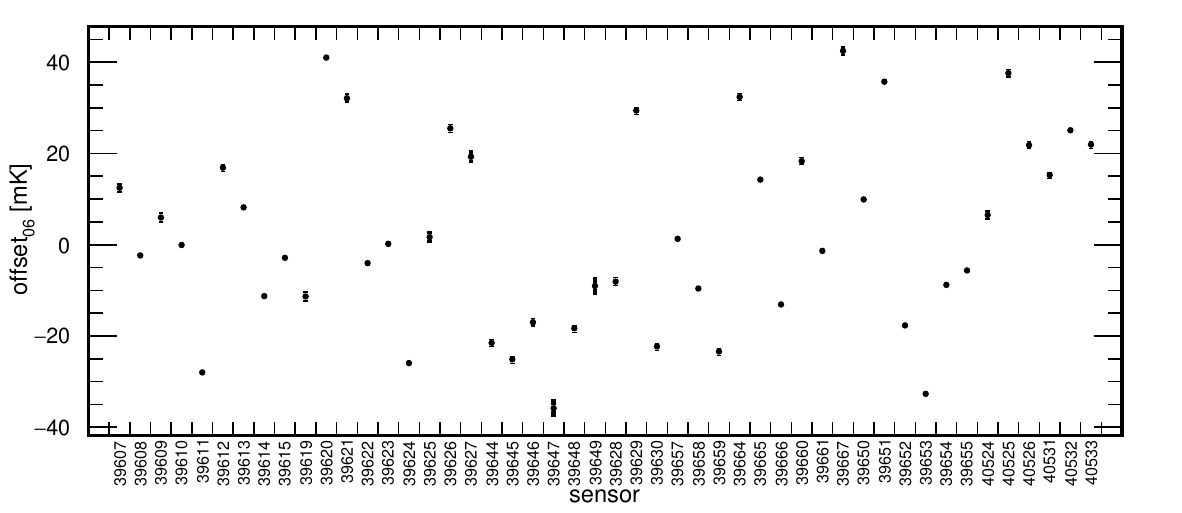}}
\caption{Offset of each sensor with respect to reference sensor 39606 using the reference method.}
\label{fig:offsets_tree_1}
\end{figure}

\noindent Fig.~\ref{fig:offsets_tree_1} shows the offset of all sensors with respect to the 39606 reference. As it can be noticed, the dispersion of the offsets is compatible with 0.1 K, the value quoted by the vendor. Fig.~\ref{fig:offsets_tree_2}-left shows the distribution of the repeatability of the computed calibration constants after applying the different corrections, showing an average value below 1 mK.

\subsubsection{Results on the tree method}
\begin{figure}[htbp]
\centering
{\includegraphics[width=0.45\textwidth]{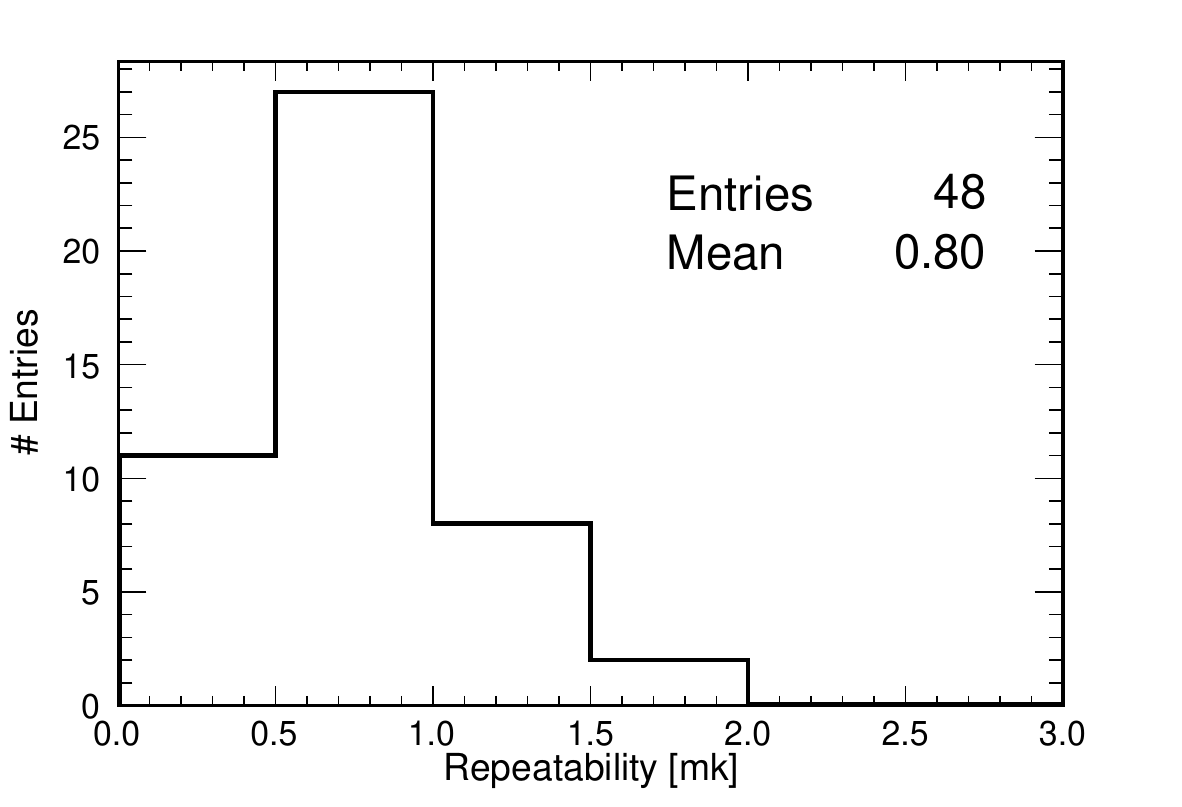}}
{\includegraphics[width=0.45\textwidth]{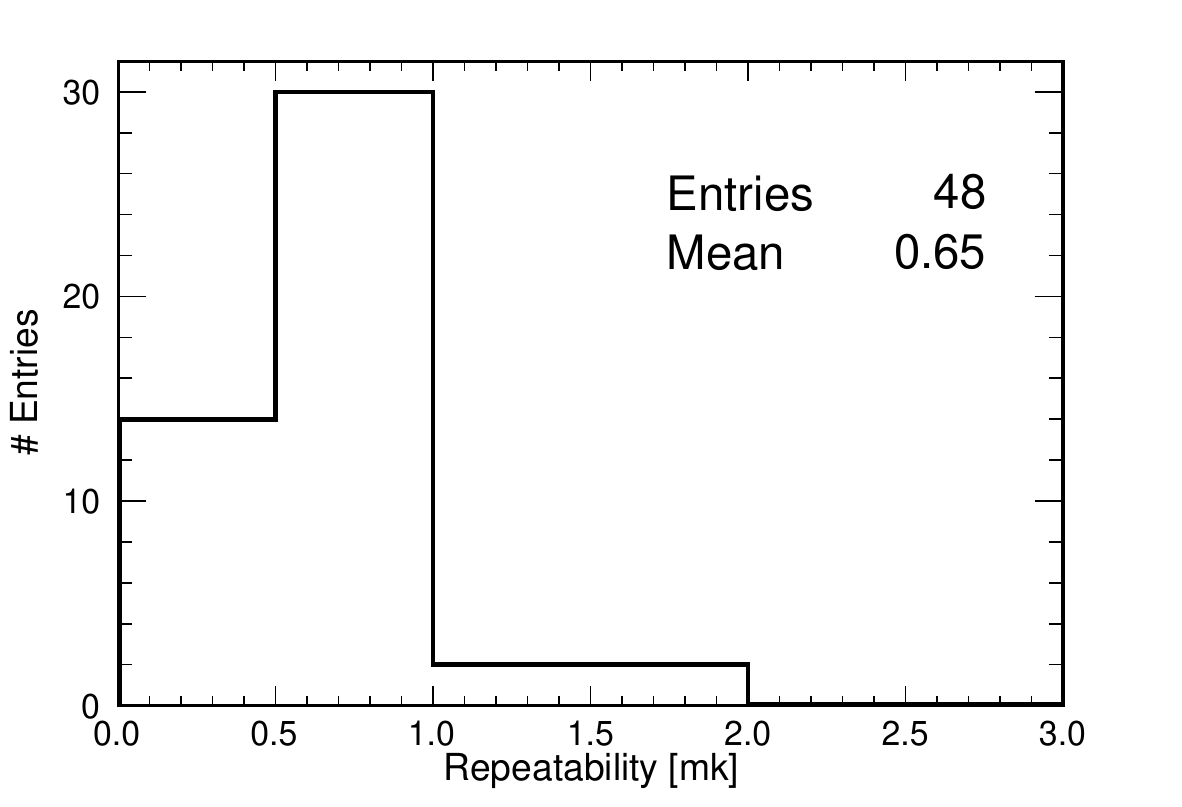}}
\caption{Left: repeatability distribution for the reference method. Right: repeatability distribution for tree method.}
\label{fig:offsets_tree_2}
\end{figure}

\noindent The offsets are computed in this case with respect to an arbitrary reference among all sensors being calibrated. Selecting as reference a sensor present in the third round (see Fig.~\ref{fi:CAL_sequence}) minimizes the number of operations required to compute the offsets, thereby reducing the associated uncertainty. Sensor 39645 was chosen as reference. Fig.~\ref{fig:offsets_tree_2}-right shows the distribution of the repeatability of the computed calibration constants, the mean of which is slightly below the one obtained for the reference method. Thus, it is confirmed that despite the higher number of intermediate sensors to relate any two sensors, the additional uncertainty introduced by the time walk correction makes the tree method superior to the reference method. Moreover, the reference method introduces a ---not yet known--- systematic error associated to the time wall correction model.

\subsubsection{Consistency cross-check and error estimation}
\label{sec:crossCheck}
\noindent The uncertainties presented are, on average, smaller than 1~mK and correspond to the standard deviation of the four independent measurements of each calibration constant, combined with the error propagation from the required corrections: time-walk correction for the reference method, addition of calibration constants for the tree method, and the uncorrected electronic residual offset common to both calibration methods. These corrections likely include unknown systematic effects impacting the final calibration constants, which need to be added to the systematic uncertainty associated to the assumption of homogeneous temperature inside the capsule. To estimate the overall precision of the calibration, the following cross-check was performed. The calibration strategy yields two independent values for each sensor’s calibration constant—one obtained from the reference method and the other from the tree method. For the sixteen sensors used in the second round of the calibration tree, these two values are linearly independent because they are derived from different sets: the reference method relies solely on first-round runs, while the tree method uses only second and third-round runs for those sensors. Therefore, these sixteen sensors provide an effective basis for estimating the calibration procedure’s precision by comparing the results obtained from the two methods.

\begin{figure}[htbp]
\centering
{\includegraphics[width=0.49\textwidth]{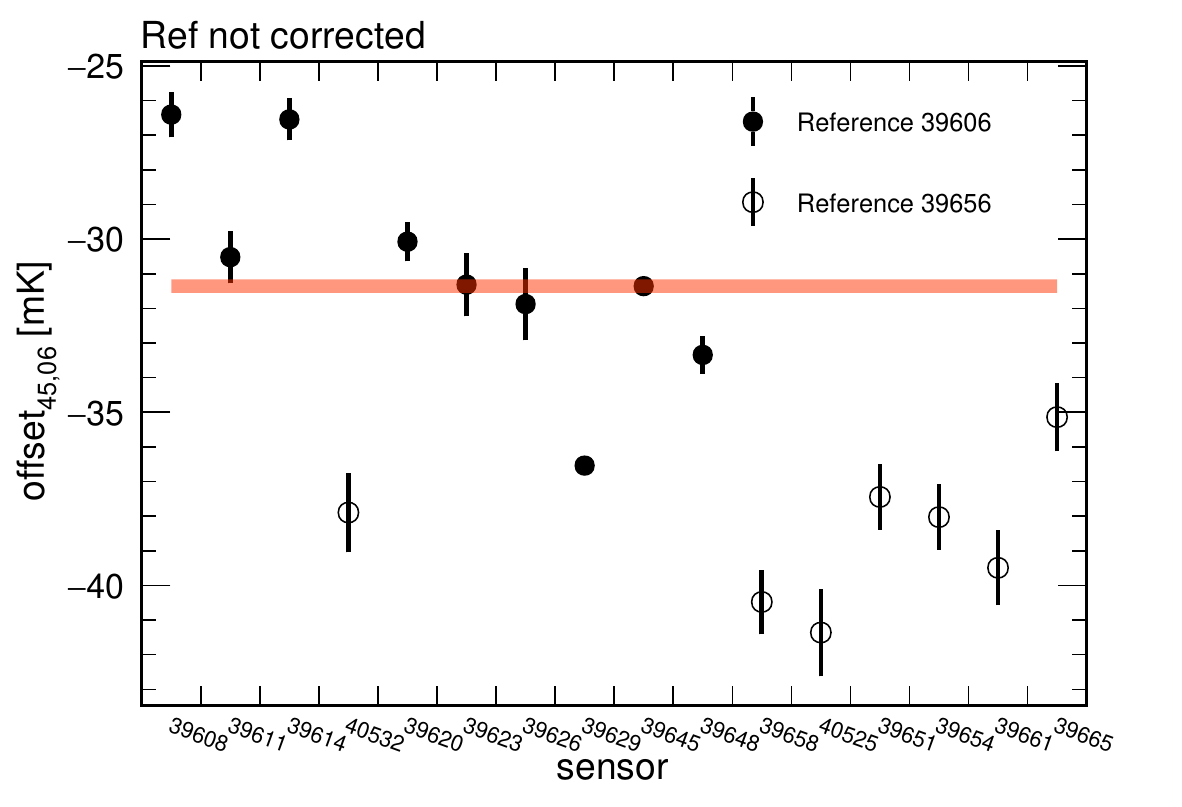}}
{\includegraphics[width=0.49\textwidth]{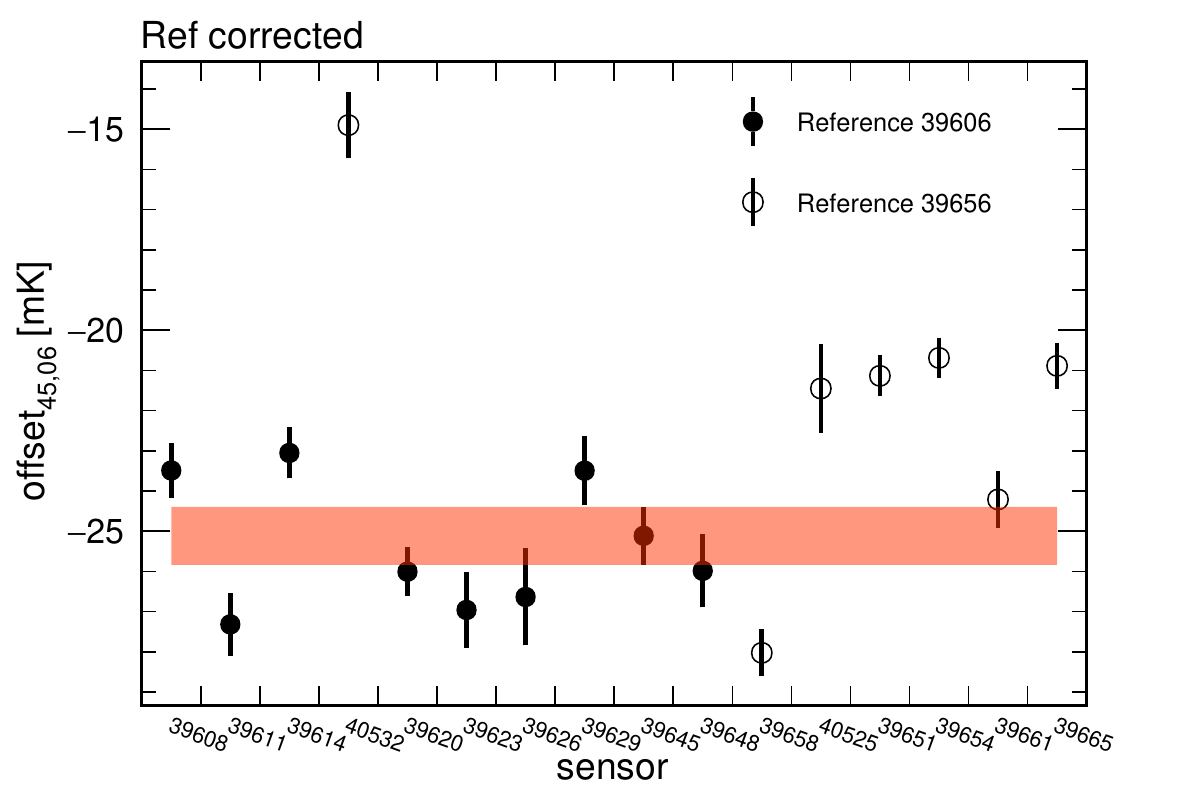}}
\caption{$\Delta T_{45,06}$ computed through all sensors of the second round of the calibration procedure. Left: not applying corrections. Right: applying corrections. A clear improvement is obtained using the corrections on the reference sensor. The red line represents the expected value of this calibration constant.}
\label{fig:crosscheck}
\end{figure}

By construction, the result of subtracting the calibration constants obtained in the two methods should be compatible with the offset between sensors 39645 and 39606, $\Delta T_{45,06} = \Delta T_{s,06}-\Delta T_{s,45} = T_{45}-T_{06}$, used as reference for the tree and reference methods respectively. This offset is found to be $-25.1 \pm 0.7$ mK by direct measurement of the two sensors (see 10th column in Fig.~\ref{fi:CAL_sequence} and \ref{fig:offsets_tree_1}).  Fig.~\ref{fig:crosscheck} shows this benchmark without (left) and with (right) time walk corrections for the 16 sensors aforementioned, probing the self-consistency of this correction. The standard deviation of this distribution, 2.4 mK, is an estimation of the quadratic sum of the total uncertainty of both calibrations. Assuming that the reference sensor method has a larger uncertainty than the tree method due to the time walk correction, an upper limit of 1.7 mK for the total error of the tree method can be assumed. In the same way, this is the inferior limit for the total error of the reference method. In both cases, these errors are less that half of what was originally required for the temperature monitoring system in DUNE FD-HD.

\section{Re-calibration after ProtoDUNE-SP operation}
\label{sec:new_calib}

\noindent After the decommissioning of ProtoDUNE-SP in 2020, the temperature sensors were disconnected from the TGM and re-calibrated multiple times (see Table \ref{tab:calib}), using both LAr and LN2 as cryogenic media. Two sensors were damaged during the decommissioning process and were not replaced until the LAr-2023 calibration campaign. The comparative analysis of those re-calibrations not only offers insights into the long-term stability of the sensors but also elucidates any potential dependencies on the choice of the cryogenic liquid. Although DUNE will use LAr, massive RTD calibration would benefit from using LN2, given its higher accessibility and lower cost.

\begin{table}[htbp]
\begin{center}
\begin{tabular}{l c c}
Date & Cryogenic liquid & \# Sensors in capsule  \\ \hline
March 2018     &   LAr    &  4  \\
February 2022  &   LN2    & 14  \\
March 2023     &   LN2    & 14  \\
July 2023      &   LAr    & 14  \\
\end{tabular}
\end{center}
\caption{Calibration runs with indication of the date, the cryogenic liquid used and the number of sensors inside the capsule.}
\label{tab:calib}
\end{table}

In this section, a description of the setup used for these new calibrations, the changes applied to the procedure as well as the results and conclusions of these calibrations are presented.

\subsection{Evolution of the calibration setup}
\label{sec:new_calib_setup}

\begin{figure}[htbp]
\centering
{\includegraphics[angle=90,width=0.3\textwidth]{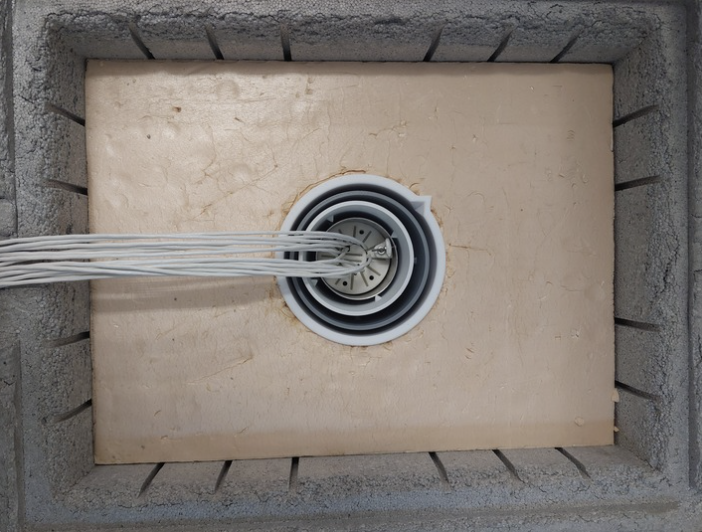}}
{\includegraphics[angle=90,width=0.297\textwidth]{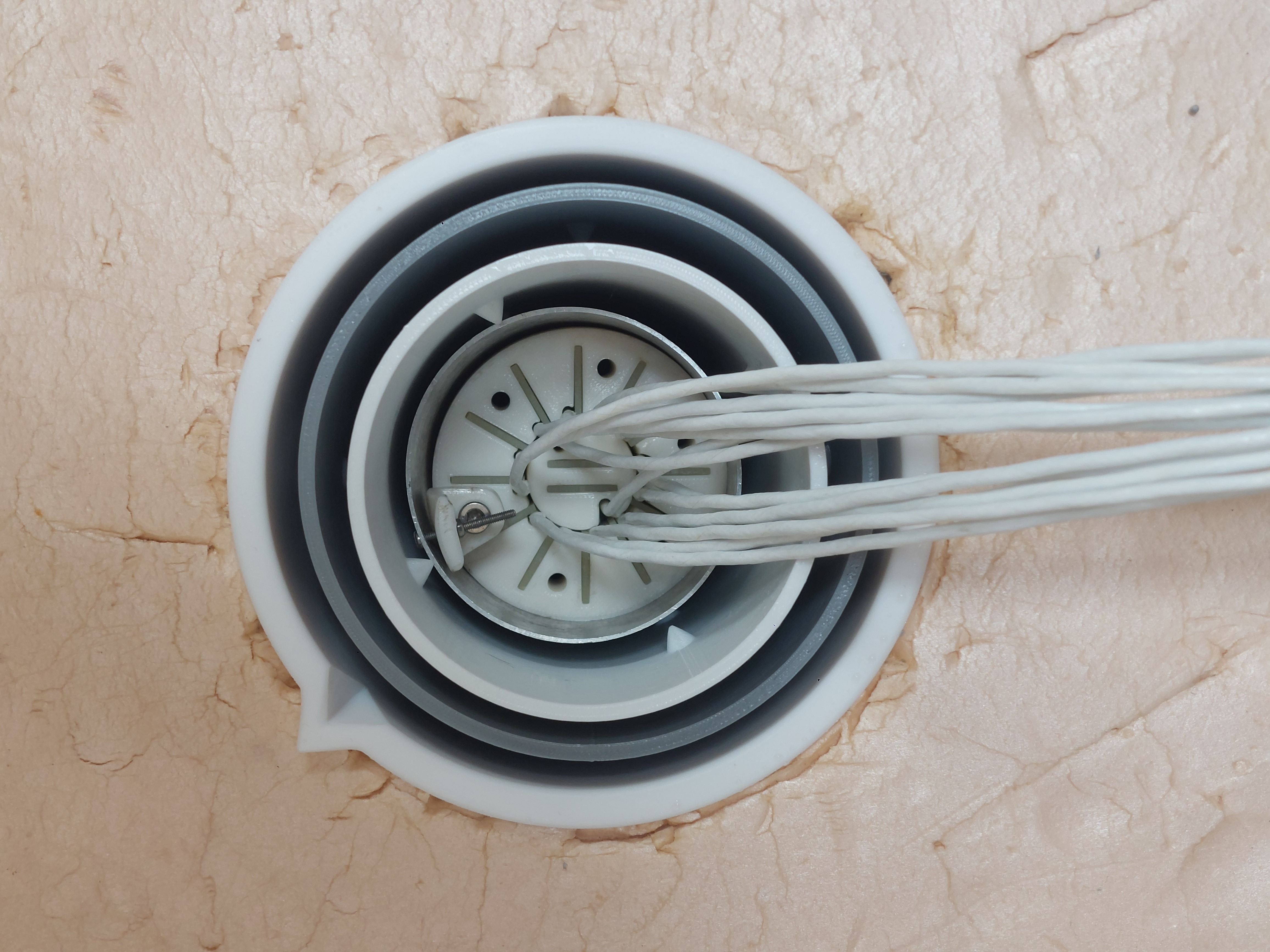}}
{\includegraphics[width=0.3\textwidth, trim=0cm 7cm 0cm 10.4cm, clip]{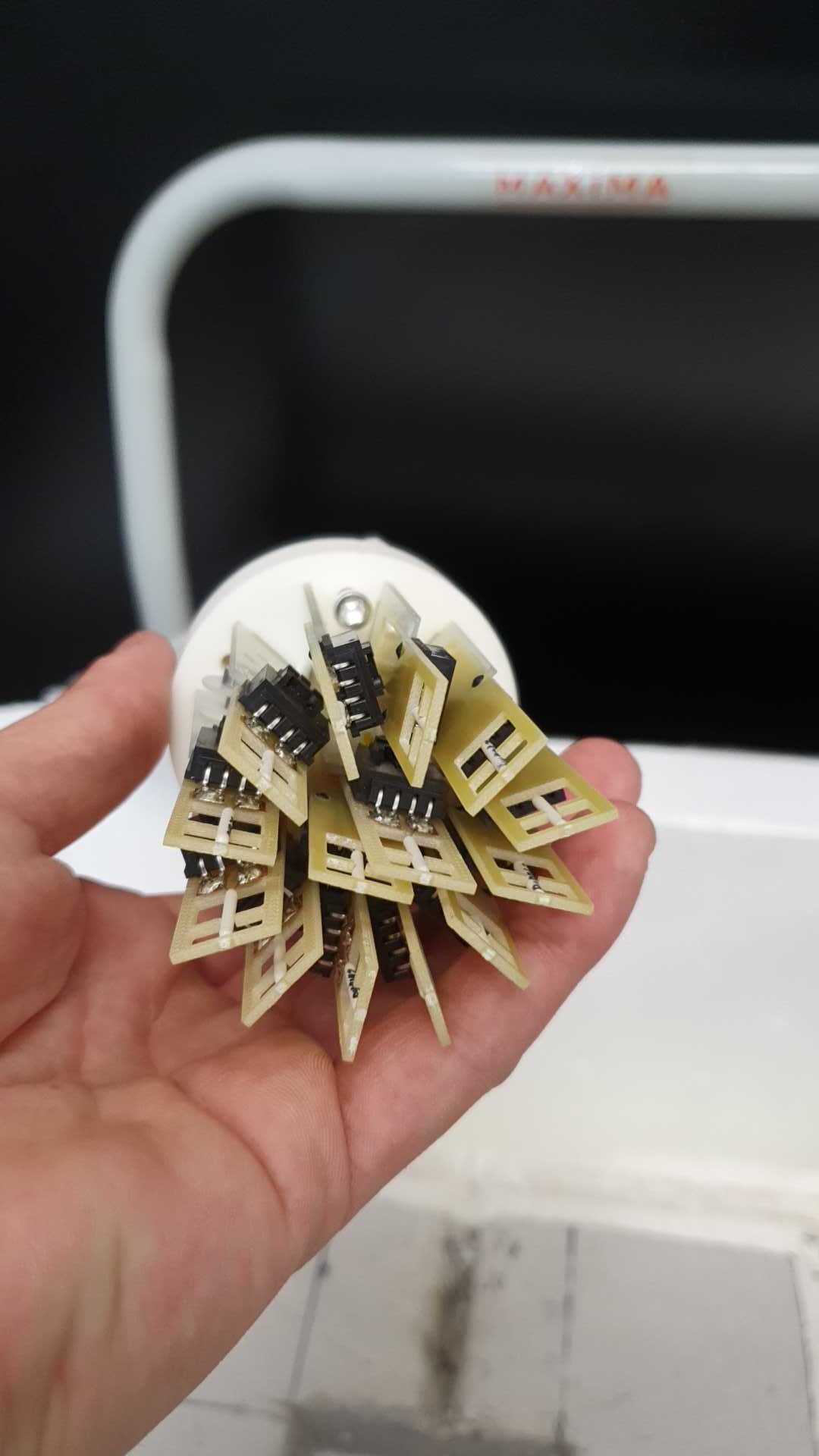}}
\caption{Left: The polystyrene box that hosts the calibration setup. Middle: Detail of the four concentric independent volumes conforming the calibration setup. Right: The 12 corona sensors plus the 2 references.}
\label{fig:newSetup}
\end{figure}

\noindent The FD-HD module will be equipped with over 500 precision sensors~\cite{dune_tdr4}, making the use of the current calibration setup impractical, as it can accommodate only four sensors simultaneously. Significant modifications have been introduced in order to increase this number to twelve, while maintaining the precision achieved during the 2018 calibration campaign. This was accomplished by positioning sensors at the same height, to avoid any potential vertical gradient, and following a cylindrical configuration, under the assumption that convection inside the capsule has rotational symmetry.
This symmetry is also kept outside the capsule in all concentric cryogenic containers, having added a fourth independent volume which should further reduce convection inside the inner capsule.
This notable advancement greatly streamlines the calibration process and serves to reduce both statistical and systematic errors associated with the procedure. Fig. \ref{fig:newSetup} shows the different elements of the new calibration setup, which are described below:

\begin{itemize}
    \item A polystyrene box with dimensions $55\times35\times30$ cm$^{3}$ and $4.5$ cm thick walls with a dedicated cover of the same material.
    \item Extruded polystyrene rectangles with a cylindrical hole of $12.5$ cm diameter and $25$ cm height, to conform the inner volume.
    \item A PTFE container with $12.5$ cm diameter, $25$ cm height and $2$ mm thick walls to fit in the hole left in the box. The approximate volume is $2$ L.
    \item A 3D printed PLA cylinder with two independent concentric volumes, placed inside the PTFE container.
    \item A cylindrical aluminum capsule to be placed in the inner volume of the PLA cylinder. It has $7$ cm diameter,  $14$ cm height and $1$ mm thick walls.
    \item A 3D printed PLA holder for 14 sensors, 12 of them forming a circle (the `corona') and the other 2 at the center, to be used as references. This support can be attached to the aluminum capsule at a fixed height. Cables are naturally extracted from the top of the assembly. A detailed view of this holder can be found in Fig. \ref{fig:newSetup}.
    \item The readout electronics, described in Sec.~\ref{sec:readout} has been retained from the previous setup.
    \item A more cost effective cable with similar performance has been used. Produced by Tempsens \cite{tempsens}, it has four twisted cables instead of two separated twisted pairs and an additional Kapton insulation layer between the shielding mesh and the four conductors. Its diameter is 2.7 mm.
\end{itemize}

\subsection{Calibration procedure}
\label{sec:new_calib_procedure}

\begin{figure}[htbp]
\centering
{\includegraphics[width=0.45\textwidth]{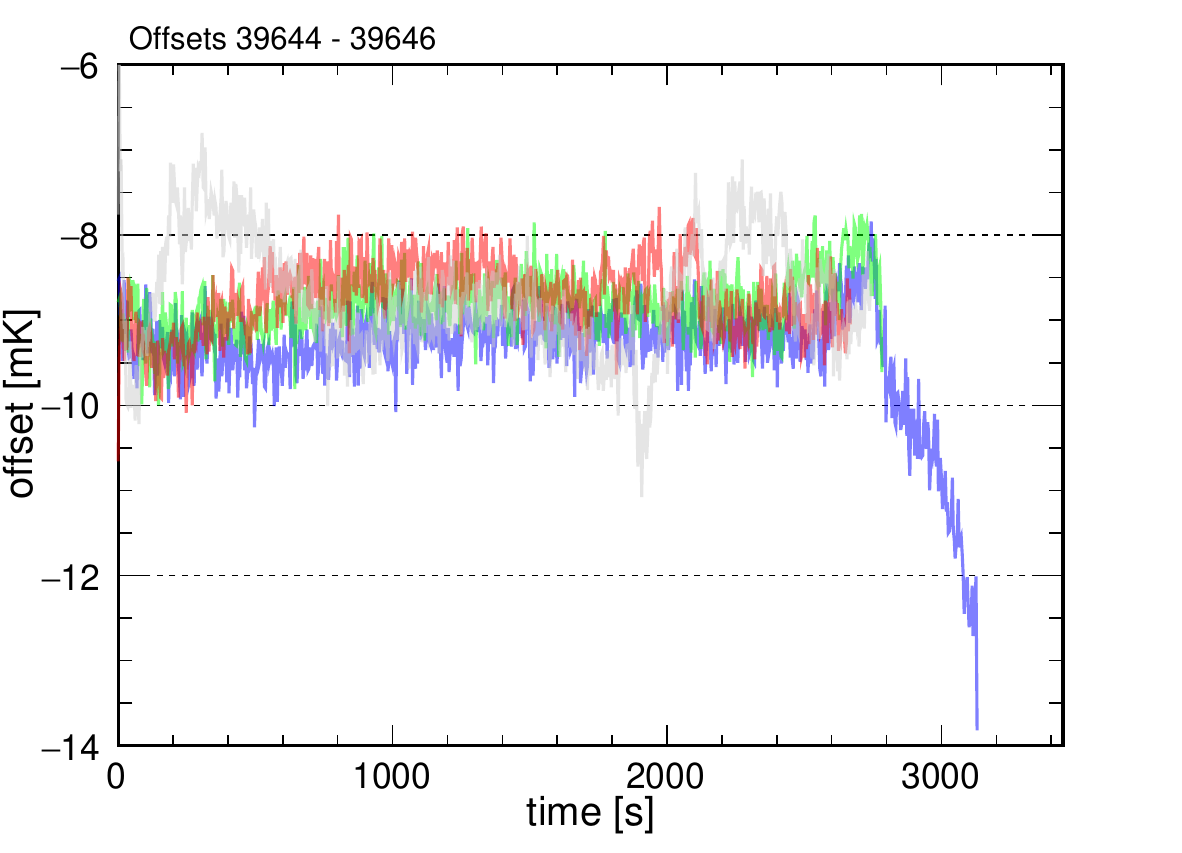}}
{\includegraphics[width=0.45\textwidth]{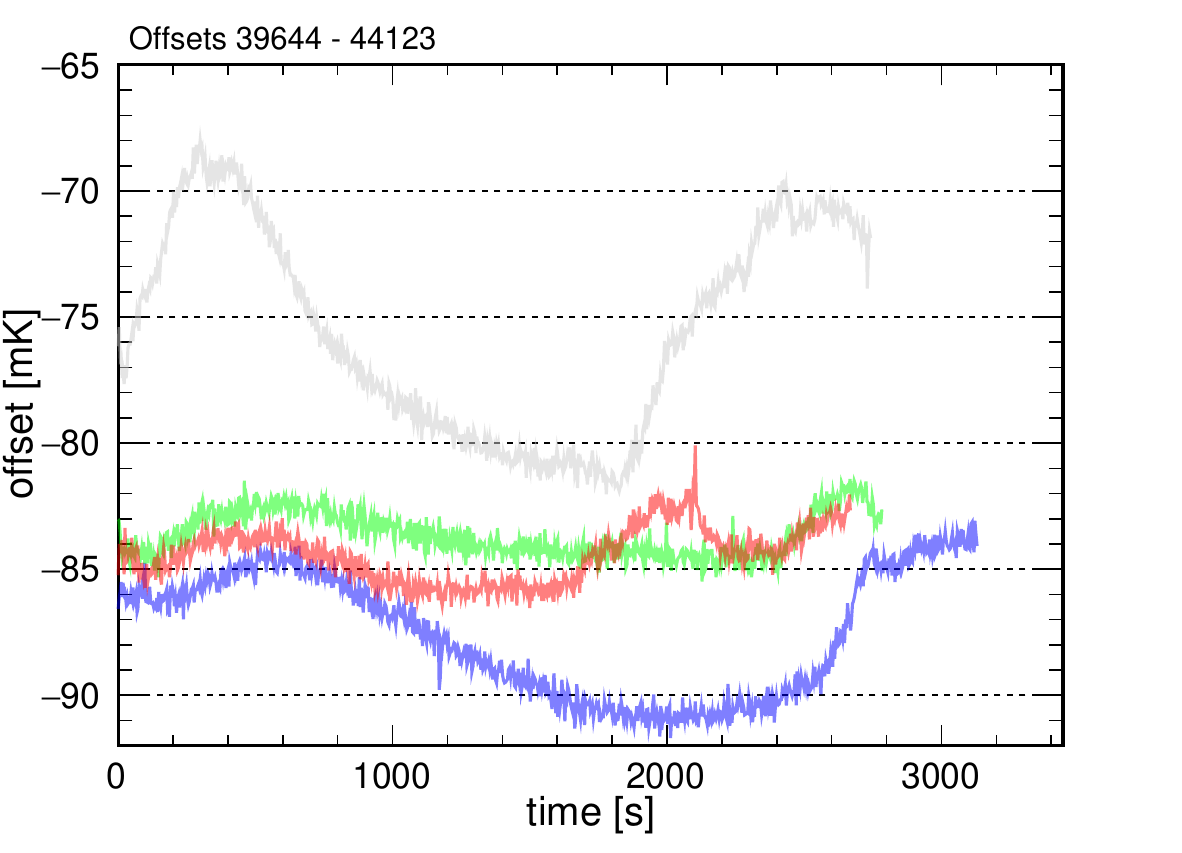}}
\caption{Left: The 4 measured offsets between two arbitrary sensors in the corona. Right: The 4 measured offsets between a sensor in the corona and one of the references, at the center.}
\label{fig:refMethodDumpingJustification}
\end{figure}

\noindent The sensor support was originally designed to accommodate two reference sensors in the center (see Fig. \ref{fig:newSetup}), enabling both the reference and tree calibration methods. However, it was soon observed that temperature variations between corona and reference sensors were significantly larger than those between any two corona sensors. This effect is attributed to the convection pattern inside the capsule, which is expected to have rotational invariance, and hence favour sensors disposed following a cylindrical symmetry. This can be observed in Fig. \ref{fig:refMethodDumpingJustification}-left, where the offset between two corona sensors is nearly constant in time in four independent measurements, and in Fig. \ref{fig:refMethodDumpingJustification}-right, where the offset between a corona and a reference sensor shows a more chaotic behaviour. Consequently, only the `tree' calibration method was considered during the calibrations after ProtoDUNE-SP decommissioning.

\begin{figure}[htbp]
\centering
{\includegraphics[width=0.9\textwidth]{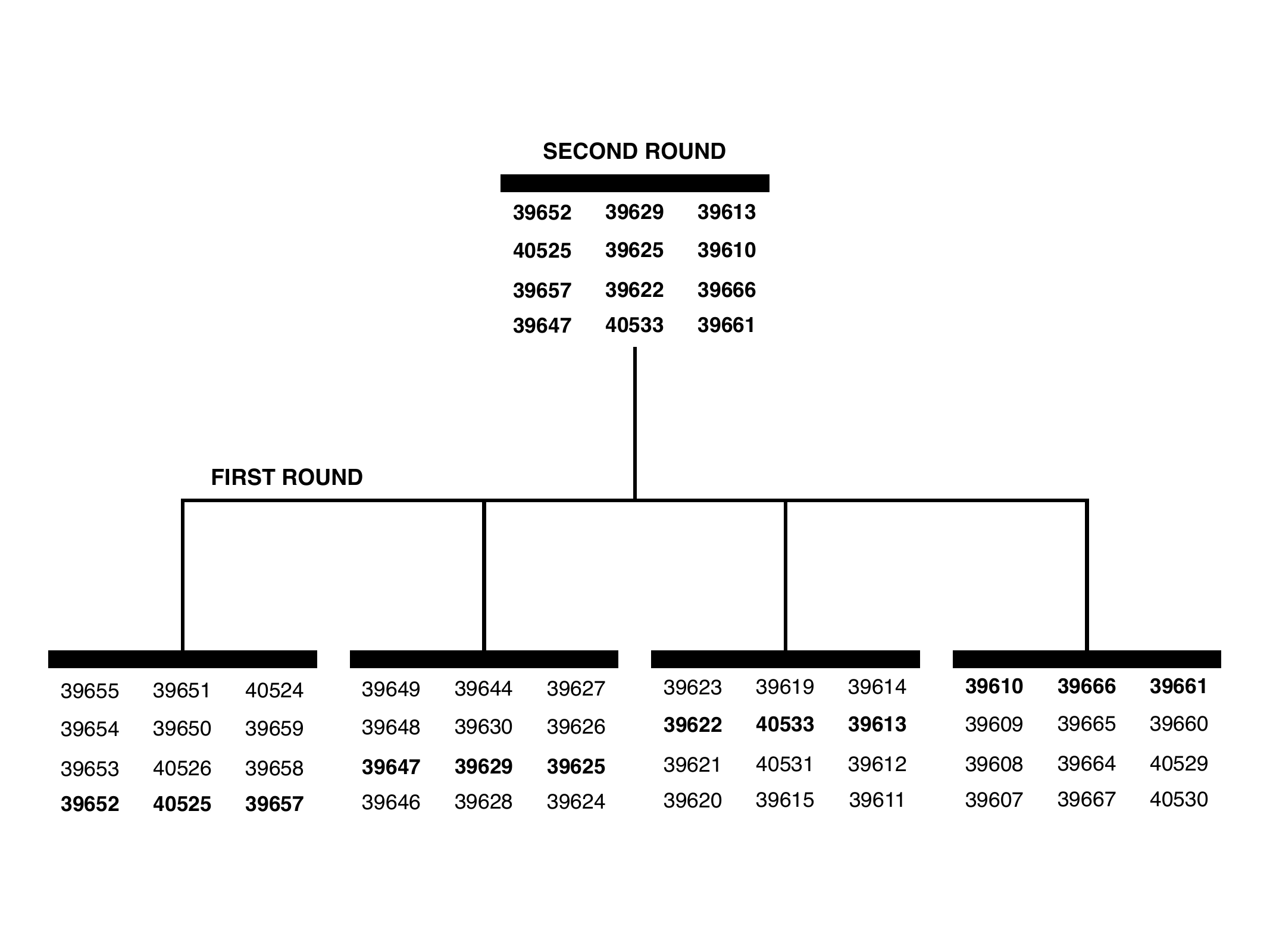}}
\caption{Schematic of the calibration sequence used with the new setup. Each of the sets contains 12 sensors, all of them placed in the corona, plus 2 references at the center. Sensors in bold in the first round are promoted to the second round.}
\label{fig:newCalibrationStrategy}
\end{figure}

The new calibration tree, shown in Fig. \ref{fig:newCalibrationStrategy}, contemplates four 12-sensors sets in the first round, and a unique second round with three promoted sensors from each of the sets in the first round. With this scheme, promoted sensors only suffer 8 baths, avoiding the necessity of a time-walk correction.

\label{sec:newCalibrationStrategy}

The experimental procedure is similar to the one presented in Sec.~\ref{sec:old_calib}. Two small variations were introduced. First, the several concentric containers are cooled down for 40 minutes before introducing the aluminium capsule for the first time in a day, slightly improving the results of the first calibration run. Second, the warm up process is accelerated using a heat gun, avoiding the use of water.

\subsection{Calibration results}
\label{sec:new_calib_results}

\begin{figure}[htbp]
    \centering
    {\includegraphics[width=0.33\textwidth]{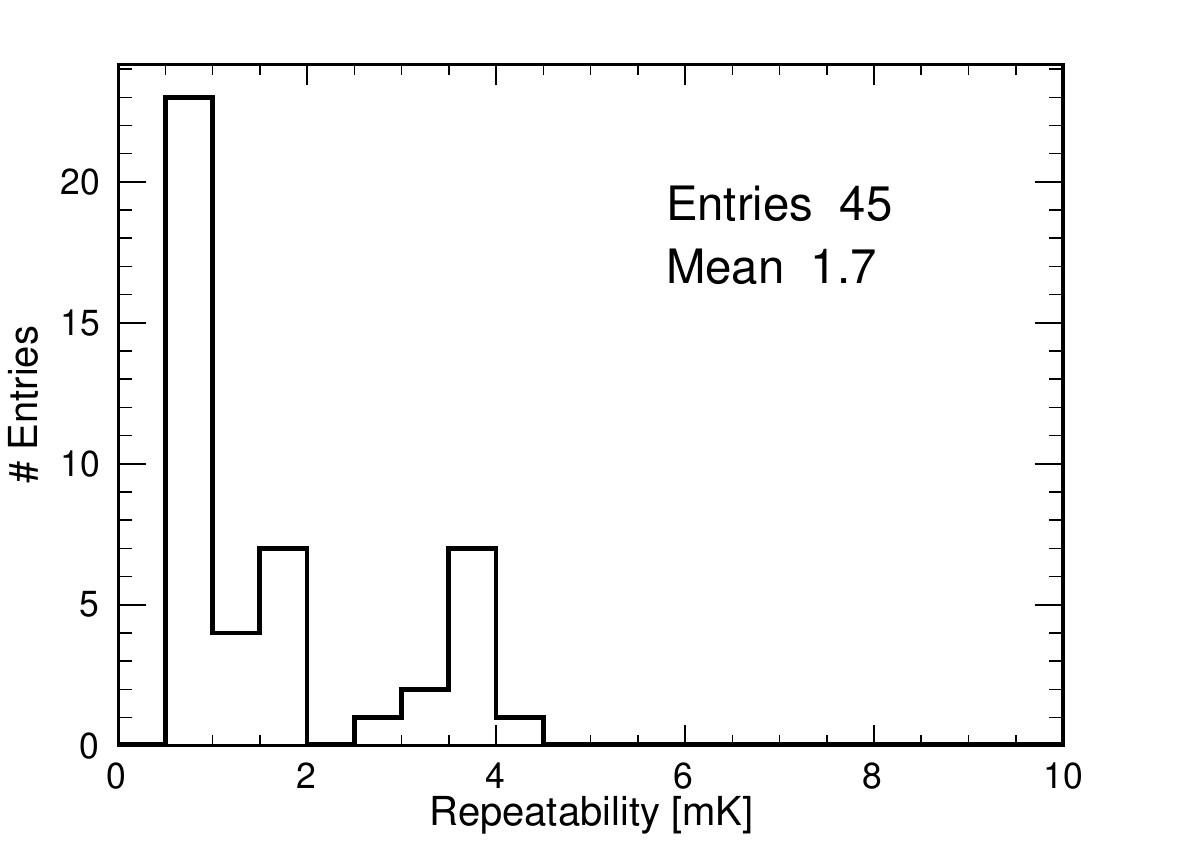}}{\includegraphics[width=0.33\textwidth]{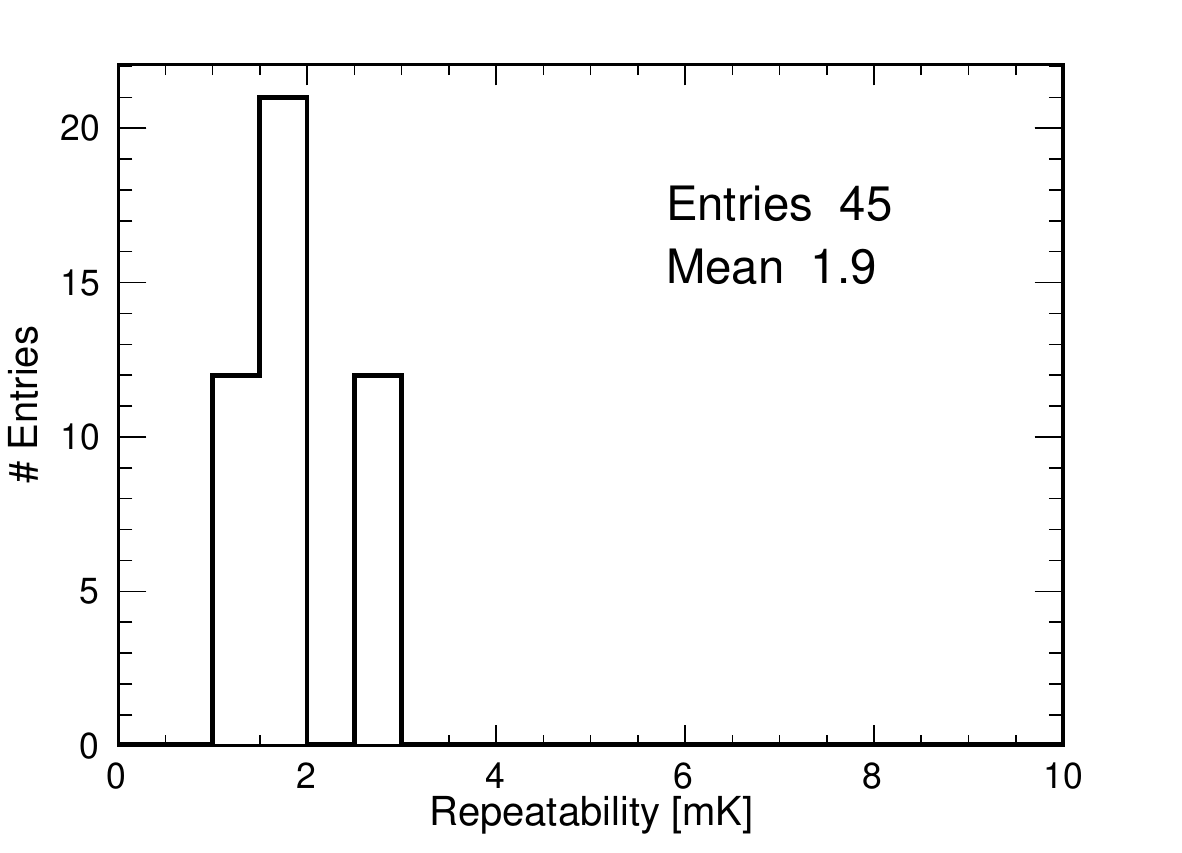}}{\includegraphics[width=0.33\textwidth]{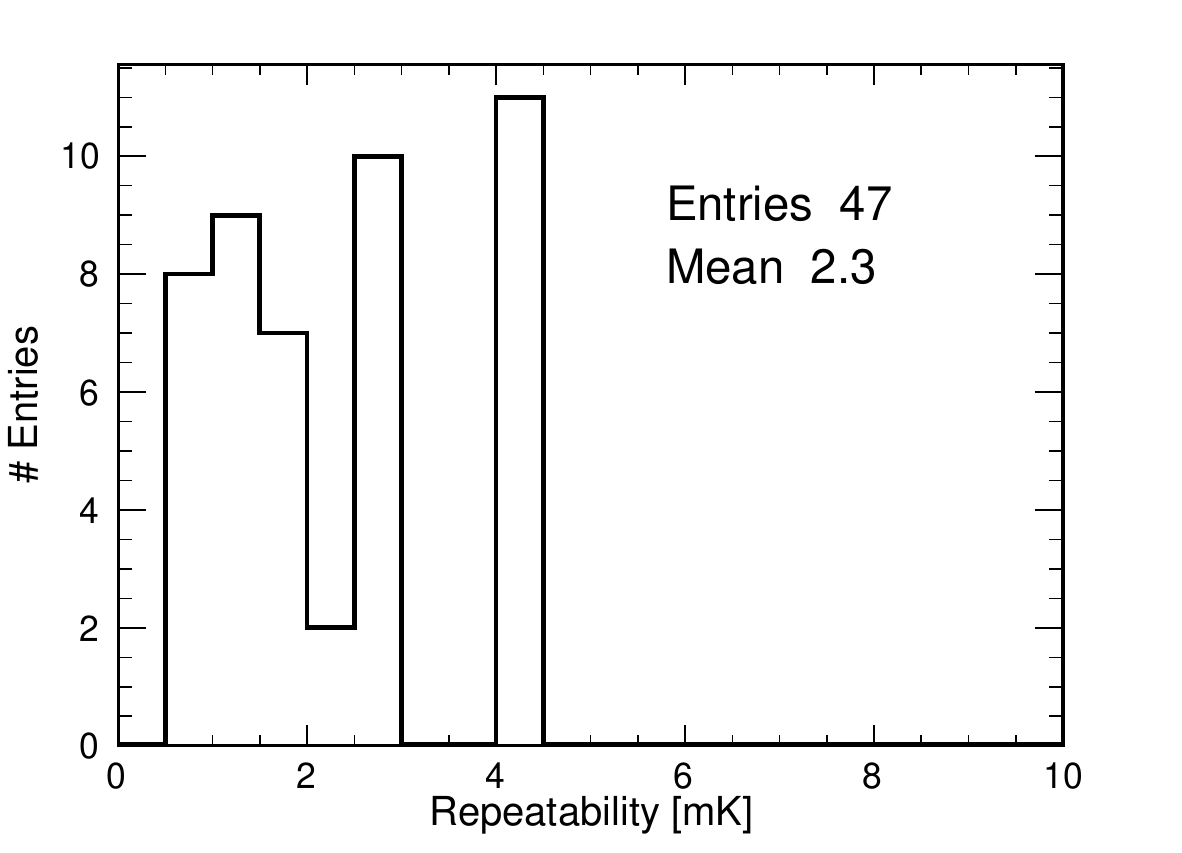}}
    \caption{Repeatability distribution for the new calibrations. Left: LN2-2022. Middle: LN2-2023. Right:LAr-2023. LAr-2023 includes two additional sensors, absent from earlier calibration campaigns, which were replaced after being damaged during the decommissioning of ProtoDUNE-SP.}
    \label{fig:newCalib_stat}
    \end{figure}

\noindent A sensor from the second round is chosen as the reference, as this minimizes the number of operations required for the non-promoted sensors. Since each set includes three promoted sensors, there are three linearly independent ways to compute the offset relative to the reference for the nine non-promoted sensors in that set. The calibration constant for these sensors is calculated as the weighted average of the three “paths”, with its uncertainty given by the standard deviation of these three values. The offset uncertainty used in the weighted average for each of the three paths is computed by quadratically summing the individual uncertainties (as in Fig.~\ref{fig:newCalib_stat}) of all terms contributing to the offset along that specific path. Sensor 40525 is adopted as the reference for the remainder of the analysis.

\begin{figure}[htbp]
\centering
{\includegraphics[width=0.65\textwidth]{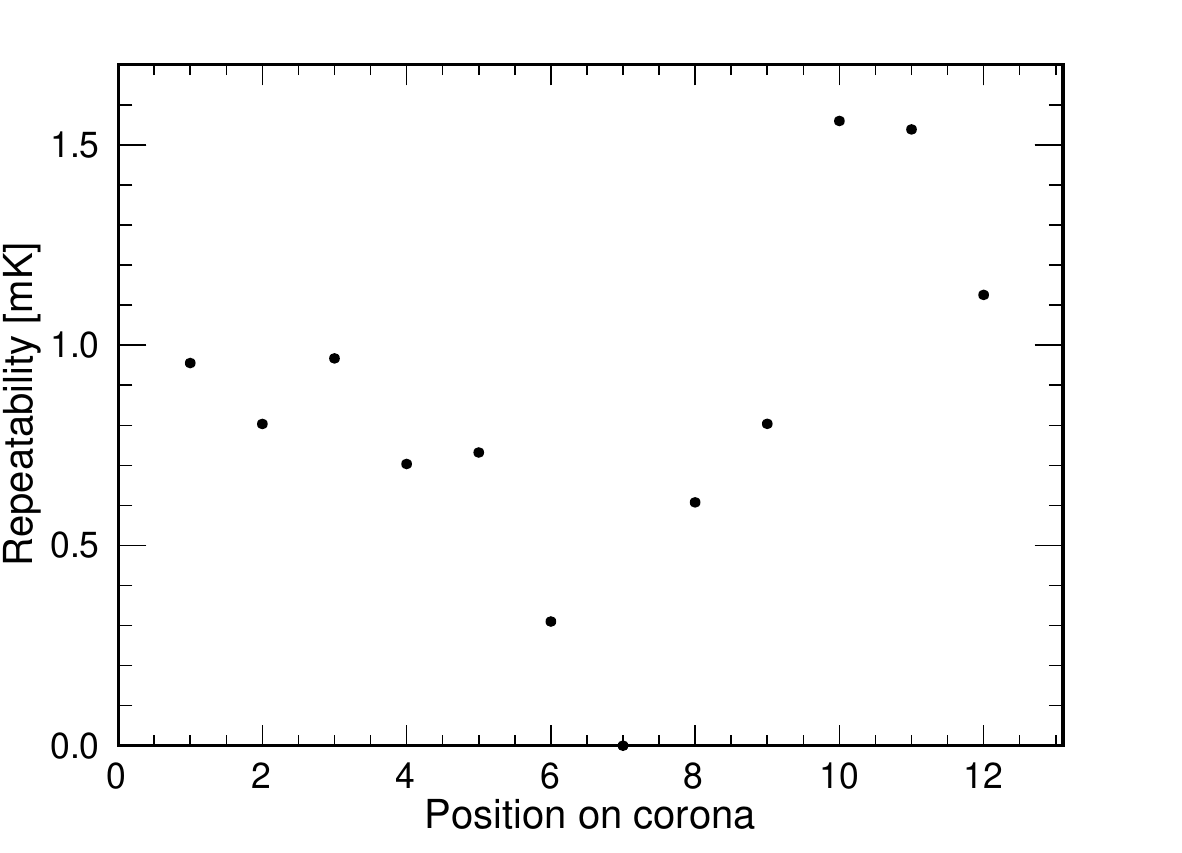}}
\caption{Repeatability as a function of position in the corona for the first set of the LAr-2023 calibration campaign. Sensor in position 7 is taken as reference.}
\label{fig:newCalib_statVsChannel}
\end{figure}

Fig.~\ref{fig:newCalib_stat} shows the distribution of the repeatability (defined in Sec.~\ref{sec:results_first_round}) for the three new calibrations mentioned in Table~\ref{tab:calib}. The values obtained are slighly worst than in the 2018 calibration campaign, which is somehow expected given the increased capsule size and the larger distance between sensors. This hypothesis is supported by Fig.~\ref{fig:newCalib_statVsChannel}, showing the repeatability as a function of the distance to the reference sensor in the corona for a particular measurement. However, there is no substantial difference between the repeatability obtained for LN2 and LAr calibrations.



\begin{figure}[htbp]
\centering
{\includegraphics[width=0.32\textwidth]{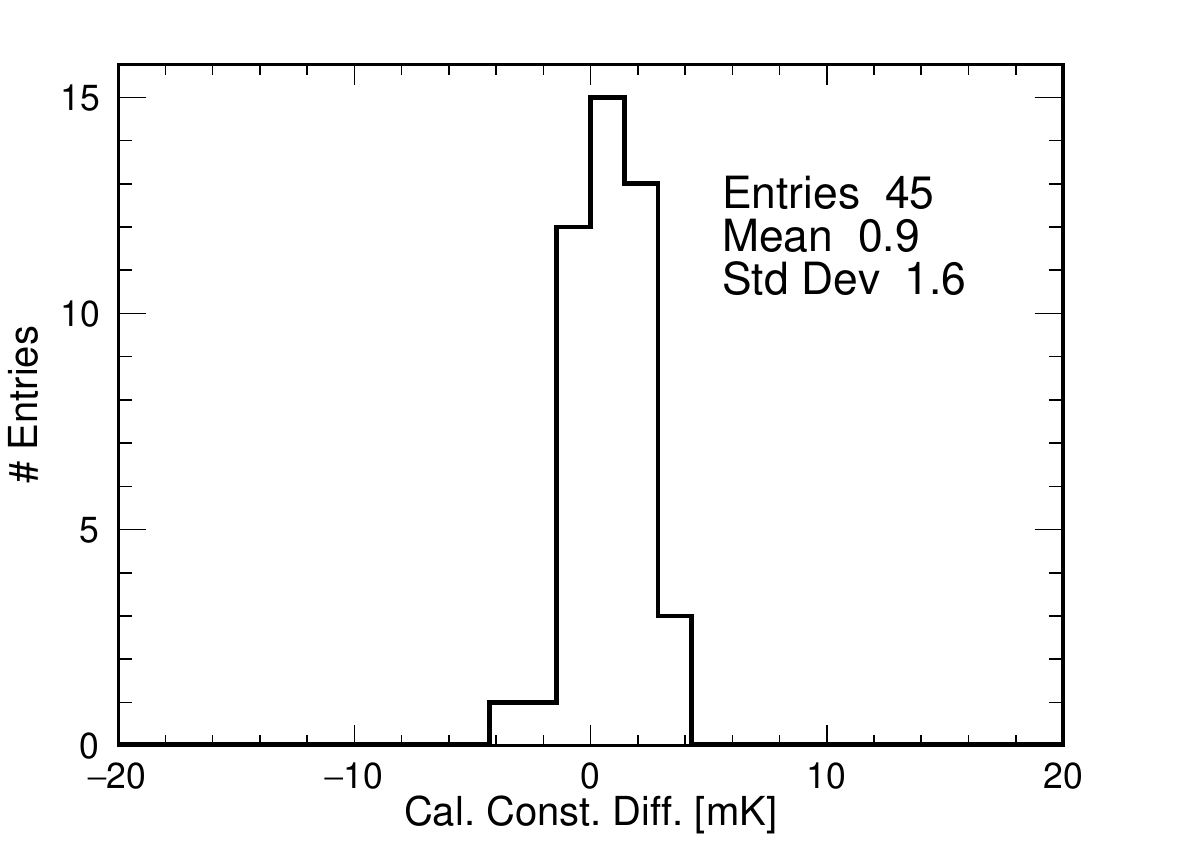}}
{\includegraphics[width=0.32\textwidth]{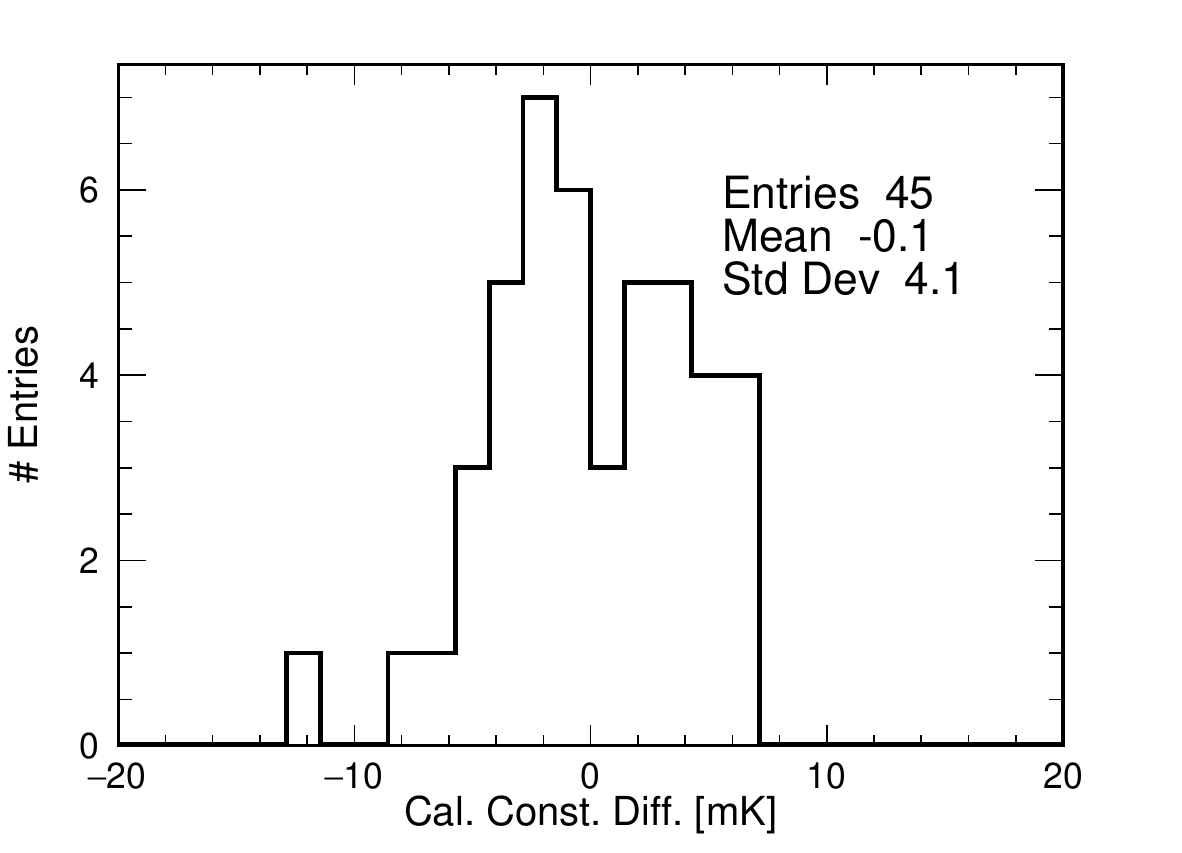}}
{\includegraphics[width=0.32\textwidth]{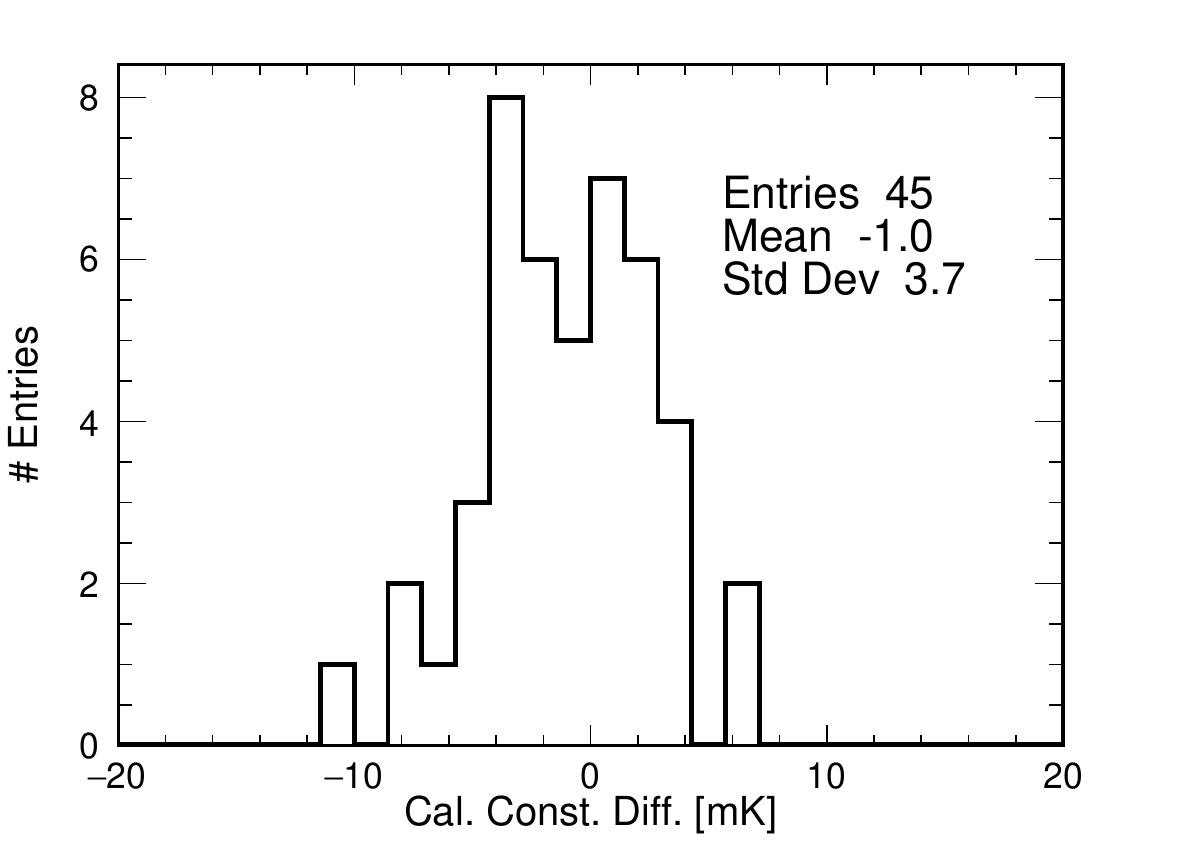}}
\caption{Difference between offsets for two calibrations. Left: LN2-2022 and LN2-2023. Middle: LN2-2022 and LAr-2023. Right:  LN2-2023 and LAr-2023.}
\label{fig:comp_newCalib}
\end{figure}

Comparison between LN2 and LAr calibrations can also be used to understand the dependence of the calibration constants on the absolute temperature. This is shown in Fig.~\ref{fig:comp_newCalib} along with other comparisons between calibration campaigns. The standard deviation of the distribution is lower when comparing two calibrations in the same liquid (LN2 in this case). However no bias is observed when comparing calibrations in two different liquids, indicating that offsets are insensitive to a 10 K variation in absolute temperature.

Ageing and long term stability have also been addressed. The left panel of Fig.~\ref{fig:comp_newCalib} shows no difference between the offsets calculated in LN2 with a delay of one year. A better understanding of this effect is achieved by comparing with the LAr-2018 calibration, as shown in the next section.

\subsection{Comparison between new and old calibrations}
\label{sec:compNewOld}

\begin{figure}[htbp]
\centering
{\includegraphics[width=0.65\textwidth]{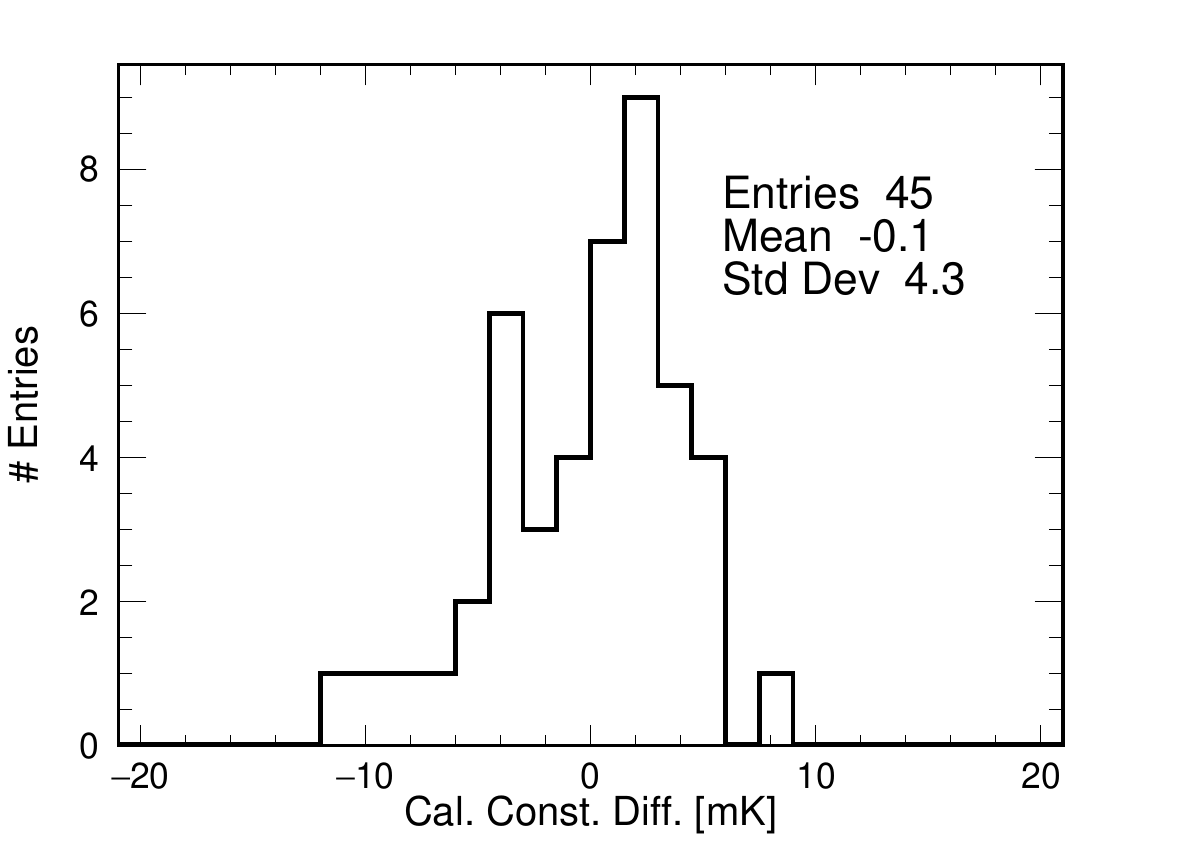}}
\caption{Difference between offsets for LAr-2018 and LAr-2023 calibration campaigns.}
\label{fig:LAr2018AllDiff}
\end{figure}

\noindent Fig. \ref{fig:LAr2018AllDiff} shows the distribution of the offset difference between the LAr-2018 and LAr-2023 calibrations. The mean of the distribution is $-0.1\pm0.6$ mK, excluding any relevant systematic drift. The standard deviation of the distribution, 4.3 mK, is only slightly larger than the ones obtained in the comparison between newer calibrations, which could be due to the unknown contribution of the uncorrected readout offsets, pointing to small or non-existing ageing effects. This can also be observed in Table~\ref{tab:calib_comparison}, summarizing the results of all possible comparisons between calibration campaigns, showing the mean and standard deviation of those comparisons. The lowest standard deviation, 1.6 mK, is obtained for the LN2-2022 to LN2-2023 combination, which is somehow expected since i) the cryogenic liquid is the same, ii) ageing should be small since there is only one year difference and iii) having use the same readout channels, readout offsets cancel out.

\begin{table}[htbp]
\begin{center}
\begin{tabular}{l c c c c}
         & LAr-2018 & LN2-2022 & LN2-2023 & LAr-2023  \\ \hline
LAr-2018 &    -     & 0.0, 3.5& 0.9, 3.9 & -0.1, 4.3 \\
LN2-2022 &    -     &    -     & 0.9, 1.6 &  -0.1, 4.1 \\
LN2-2023 &    -     &    -     &    -     & -1.0, 3.7 \\
LAr-2023 &    -     &    -     &    -     &     -     \\
\end{tabular}
\end{center}
\caption{Mean and standard deviation of the difference between the calibration constants obtained in two calibration campaigns.}
\label{tab:calib_comparison}
\end{table}

\noindent The difference between the constants obtained in two calibration campaigns can be used to estimate the total calibration error. Since ProtoDUNE Horizontal Drift (ProtoDUNE-HD), the next iteration of the ProtoDUNE-SP detector, and DUNE will use LAr, the best estimation of the error would come from a LAr to LAr comparison close in time to the actual detector running. Being this combination not available, three other combinations can be used. The comparison between the LN2 calibrations from 2022 and 2023 yields a standard deviation of 1.6~mK, which should represent the quadratic sum of the individual calibration errors. Assuming equal errors for both campaigns, this implies a single calibration error of approximately 1.13~mK. In this case, readout channel offsets cancel out, although real fluctuations in those offsets still contribute to the estimated single calibration error. The comparison between LAr calibrations from 2018 and 2023 provides further insight, with a standard deviation of 4.3~mK, corresponding to a single calibration error of about 3.0~mK. However, this value includes uncorrected readout offsets as well as potential effects from sensor ageing. Finally, comparing the new LAr and LN2 calibration campaigns avoids issues related to readout offsets and ageing but may be affected by the differences in cryogenic liquids. These comparisons yield a similar standard deviation around 3.8~mK, implying a single calibration error of approximately 2.7~mK. This value thus establishes an upper limit on the LAr calibration error. In summary, the individual calibration error in LAr is estimated to lie between 1.6 and 3.0~mK. This range is consistent with an independent estimate of 1.7~mK obtained using a different method for the LAr-2018 calibration (see Sec.~\ref{sec:crossCheck}).

\section{Conclusions}
\label{conclusions}

\noindent The DUNE experiment will require the construction and operation of the largest cryostats ever used in a particle physics experiment. This makes the continuous measurement of temperature gradients in liquid argon crucial for monitoring the stability of the cryogenics system and for detector calibration. R\&D on the calibration of RTD probes started in 2017, leading to promising results for sensors installed in the DUNE prototype at CERN.

The first setup proved the viability of the method, obtaining a calibration error of 1.7 mK. A key component was the readout electronics, with an intrinsic resolution better than 0.5 mK in the comparison between two channels. The mechanics was also crucial, with several insulation layers consisting of independent concentric volumes, ensuring minimal convection in the inner volume. Sensors were contained in an aluminium capsule, enabling slow cool-down and warm-up processes, found to be fundamental to guarantee the integrity of the sensors and to minimize the effect of ageing.

The calibration system was later enhanced to accommodate the large-scale calibration required for the DUNE detectors. The capacity of the inner capsule was increased from 4 to 14 sensors, while improvements were made to the insulation and symmetry of the system to minimize temperature differences between sensors. The new system has slightly worst repeatability for sensors in the same set, but reduces the statistical and systematic errors associated to the calibration tree, needed to relate any two sensors in different calibration sets. Calibration with the new setup has achieved a precision in the range of 1.6-3.0 mK, substantially better than the 5 mK DUNE requirement.

Another difference between the new and old calibrations is the use of a different cryogenic liquid. While DUNE will use LAr, LN2 is cheaper, simplifying the process of massive calibration for DUNE detectors. The 10 K difference between those liquids has a minimal effect on the calibration constants.

A comparison of the four calibration campaigns has provided valuable insights into aging effects, with no evidence of RTD aging observed over a five-year period. This highlights the stability and reliability of the PT-102-based system.
\section{Acknowledgments}
\noindent The present research has been supported and partially funded by Conselleria d’Innovació, Universitats, Ciència i Societat Digital, by the Fundación `La Caixa', and by MICINN. The authors would would also like to thank all the people involved in the installation, commissioning and operation of ProtoDUNE-SP, specially to Stephen Ponders. 

\bibliographystyle{elsarticle-num}
\bibliography{bibliography}

\end{document}